\documentclass{aa}
\usepackage{graphicx}
\usepackage{natbib}
\bibpunct{(}{)}{;}{a}{}{,}
\usepackage{epsfig}
\usepackage{txfonts}
\usepackage{supertabular} 
\begin{document}

\title{Fluxes in H$\alpha$ and Ca~II H and K for a sample of Southern stars}
\title{ H$\alpha$ and the Ca~II H and K lines as activity proxies for late-type	 stars}

 
\author{C. Cincunegui\inst{1,2}$^{,\star}$, 
R. F. D\'{\i}az\inst{1,2}$^{,\star}$ \and P. J. D. Mauas\inst{1,3}$^{,}$
\thanks{Visiting Astronomers, Complejo Astron\'omico  
El Leoncito operated under agreement between the Consejo Nacional 
de Investigaciones Cient\'{\i}ficas y T\'ecnicas de la Rep\'ublica Argentina 
and the National Universities of La Plata, C\'ordoba and San Juan.}}
 
\offprints{C. Cincunegui, \email{carolina@iafe.uba.ar}}
 
\institute{Instituto de Astronom\'{\i}a y F\'{\i}sica del Espacio, CONICET-UBA,
C.C. 67 Suc. 28 (1428) Buenos Aires, Argentina \and Fellow of the CONICET \and
Member of the Carrera del Investigador Cient\'{\i}fico, CONICET}
 
\authorrunning{C. Cincunegui, R. F. D\'iaz \and P. J. D. Mauas}

\date{Received <date>; accepted 
<date>} 
\abstract
{The main chromospheric activity indicator is the $S$ index, which is esentially the ratio of the flux 
in the core of the Ca II H and K lines to the continuum nearby, and is well 
studied basically for stars from F to K. Another usual chromospheric proxy
is the H$\alpha$ line, which is beleived to be tightly correlated with the Ca~II index.}
{In this work we characterize both chromospheric activity indicators, one 
associated with the H and K Ca~II lines and the other with H$\alpha$, for the 
whole range of late type stars, from F to M.}
{We present periodical medium-resolution echelle observations 
covering the complete visual range, which were taken at the CASLEO Argentinean 
Observatory. These observations are distributed along 7 years. We use a total 
of 917 flux-calibrated spectra for 109 stars which range from F6 to M5. We 
statistically study these two indicators for stars of different activity 
levels and spectral types.}
{We directly derive the conversion factor which translate the known $S$ index 
to flux in the Ca~II cores, and extend its calibration to a wider spectral 
range. We investigate the relation between the activity measurements in the 
calcium and hydrogen lines, and found that the usual correlation observed is 
basically the product of the dependence of each flux with stellar colour, and 
not the product of similar activity phenomena.}
{} 

 
\maketitle
 
\newcommand{\hal}{\ensuremath{$H$\alpha}}
\newcommand{\ca}{\ensuremath{$Ca$ $~II$ }}
\newcommand{\hel}{\ensuremath{$D$_3}}
\def\teff{\hbox{$T_{\mathrm{eff}}$}}
\def\deg{\hbox{$^\circ$}}
\def\sun{\ensuremath{\odot}}
\newcommand{\prot}{\ensuremath{P_\mathrm{{rot}}}}

\newcommand{\1}{$^a$}
\newcommand{\2}{$^b$}
\newcommand{\3}{$^c$}
\newcommand{\4}{$^d$}
\newcommand{\5}{$^e$}
\newcommand{\6}{$^f$}
\newcommand{\7}{$^g$}
\newcommand{\8}{$^h$}
\newcommand{\9}{$^i$}
\newcommand{\0}{$\circ$}
\newcommand{\z}{$^{(j)}$}
\newcommand{\zz}{$^{(k)}$}

\newcommand{\metal}{[Fe/H]}
\newcommand{\gravity}{$\log g$}
\newcommand{\vsini}{$v \sin i$}
\newcommand{\col}{$B\!-\!V$}
%

\section{Introduction}

In the Sun, the magnetic fields which are generated by turbulence in the outer 
convection zone and penetrate the solar atmosphere cause a very broad range of 
phenomena, such as sunspots, plages, active regions, flares, etc. The term 
stellar activity refers to similar features that occur in late-type stars, 
i.e., in stars which also have an outer convection zone. 
It is generally accepted that magnetic activity in late-type stars is the 
product of an $\alpha \Omega$ dynamo, which results from the action of 
differential rotation at the tachocline (the interface between the convective 
envelope and the radiative core). 

Therefore, the nature of stellar activity is closely related to the existence 
and depth of an outer convection zone. Since this depth depends on spectral type 
--from F stars which have shallow convection zones to middle M stars which are 
totally convective--, it is of special interest to characterize the stellar 
activity in stars of different spectral types.

To date, the usual indicator of chromospheric activity is 
the well known $S$ index, esentially the ratio of the flux in the core of the Ca 
II H and K lines to the continuum nearby \citep{1978PASP...90..267V}. 
This index has been defined at the Mount Wilson Observatory, were an extensive 
database of stellar activity has been built over the last four decades. 
However, these observations are mainly concentrated on stars ranging from F to K 
\citep[see, for example,][]{1995ApJ...438..269B}, due to the long exposure times 
needed to observe the Ca II lines in the red and faint M stars. For this 
reason, the $S$ index is poorly characterized for these stars. 
In spite of that, many attempts have been made to study activity using 
the \ca\ lines \citep[see, for example][]{1989A&A...213..261P, 
1991ApJS...76..383D, 1996AJ....111..439H, 2004ApJS..152..261W}.

Nevertheless, as we pointed out, the H and K lines are not the most adequate 
activity indicator for K and M stars, due both to their red color and to their 
intrinsically faint luminosity. Another well studied activity indicator is the 
\hal\ line \citep[see, for example][]{1991A&A...251..199P, 1995A&A...294..165M}. 
In fact, it has been claimed that it does exist a strong correlation between 
\hal\ and the \ca\ lines \citep{1995A&A...294..165M, 1990ApJS...72..191S, 
1990ApJS...74..891R, 1989ApJ...345..536G}. 

In this work we characterize both chromospheric indicators for a sample of 
more than 100 Southern stars. We use more than 900 flux-calibrated spectra which 
include the full spectral range between both proxies. 

The paper is organized as follows. In Section~\ref{sec:stars} we explain how we 
choose the stellar sample and present the observations, and describe briefly the method 
employed to calibrate them. In Section~\ref{sec:Ca} we 
analyze the \ca\ lines and its continuum nearby, and the \hal\ line and its 
continuum are studied in Section~\ref{sec:Hal}. Since in our spectra both fluxes 
are measured simultaneously, we are able to study in great 
detail the correlation between the \ca\ lines and \hal . We perform this 
analysis also in Section 4. Finally, in Section~\ref{sec:sumario} we 
summarize our results.


\section{The observations}\label{sec:stars}

\subsection{The sample of stars}

In this work we study a sample of 109 Southern stars chosen in order to 
cover the whole range of depths of the convective zone ($0.45 \le$ \col\ 
$\le 1.81$), from the onset of the outer convective layer to fully convective 
stars. They were also selected to include different levels of chromospheric 
activity, from stars known to have non varying records to flare stars.

In Table~\ref{tab:stars} (at the end of the paper) we list the stars included in 
the sample. For each star 
we give the HD and GJ/GL numbers and the star name in the three leftmost 
columns. In the fourth column we indicate which stars are expected to be 
particularly active, either because they are dMe stars or because they belong to 
the RS~CVn or the BY~Dra types. In that column we also specify which stars are 
chosen as non-variable by \citet{1996AJ....111..439H}, based on the long-term 
observations of the Mount Wilson database of the \ca\ lines. Also indicated are 
the stars for which planetary systems have been reported.

In the fifth to seventh columns we list the star's spectral type, its $V$ 
magnitude and \col\ colours, taken from~\citet{1997A&A...323L..49P}, except for 
the ones indicated. Also from the same catalogue are the parallaxes (in mas), 
in the eigth column. The gravity and metallicity in the next two columns are 
taken from \citet{1997A&AS..124..299C,2001A&A...373..159C}. The eleventh column 
lists the rotational periods, either measured or estimated from \vsini , as 
stated. In column number twelve  
we include the number of observations of each star used in this work.

Finally, in the last five columns we give the mean activity indexes 
$\langle S \rangle$ (in Mount Wilson units), $\langle F_{\mathrm{HK}} \rangle$ 
and $\langle F_{\mathrm{H}\alpha} \rangle$, as defined in Sections~\ref{sec:S} 
and \ref{sec:FHa}, and the parameters of the linear regresions of 
Section~\ref{sec:AyS}.


\subsection{The observations and their processing}

Our observations were made at the 2.15~m  telescope of the Complejo 
Astron\'omico El Leoncito (CASLEO), which is located at 2552~m above sea level, 
in the argentinian Andes. The medium-resolution echelle spectra were obtained 
with a REOSC spectrograph designed to work between 3500 and 7500~\AA\ and a 
1024~$\times$~1024 pixel TEK CCD as detector. The maximum wavelength range 
of our observations is from 3860 to 6690~\AA\ 
($R=\lambda/\delta\lambda \approx 26\,400$).

We calibrated in flux all the spectra following the procedure described in 
\citet{2004A&A...414..699C}. The method makes use of long-slit spectra of the 
same target stars. We calibrated this low-resolution spectra with 
spectrophotometric standard stars by the usual procedure. Then, for each 
echelle spectrum, we flux-calibrated it using the long-slit spectrum of the same 
star as if it were a standard star. In \citet{2004A&A...414..699C} 
there are various examples of these spectra, together with a discussion about 
the errors involved in the calibration procedure.

We began our observing program in 1999, and at present we carry on four to five 
observing runs per year. 
In this work we use a total of 917 spectra for 109 stars.


\section{\ca\ index}\label{sec:Ca}

The index most extensively used for chromospheric activity studies is the well 
known Mount Wilson $S$ index. Nevertheless, it is not well characterized for the 
less masive stars. As we explained, because of the long exposure times needed to 
observe these stars, usually they are not included in the stellar activity 
databases. In this section we study the behaviour of the 
$S$ index in the whole range of stars with a radiative core and a convective 
outer layer, i.e., from mid F to mid M.

\begin{figure*}[t]
\centering 
\includegraphics[width=17cm]{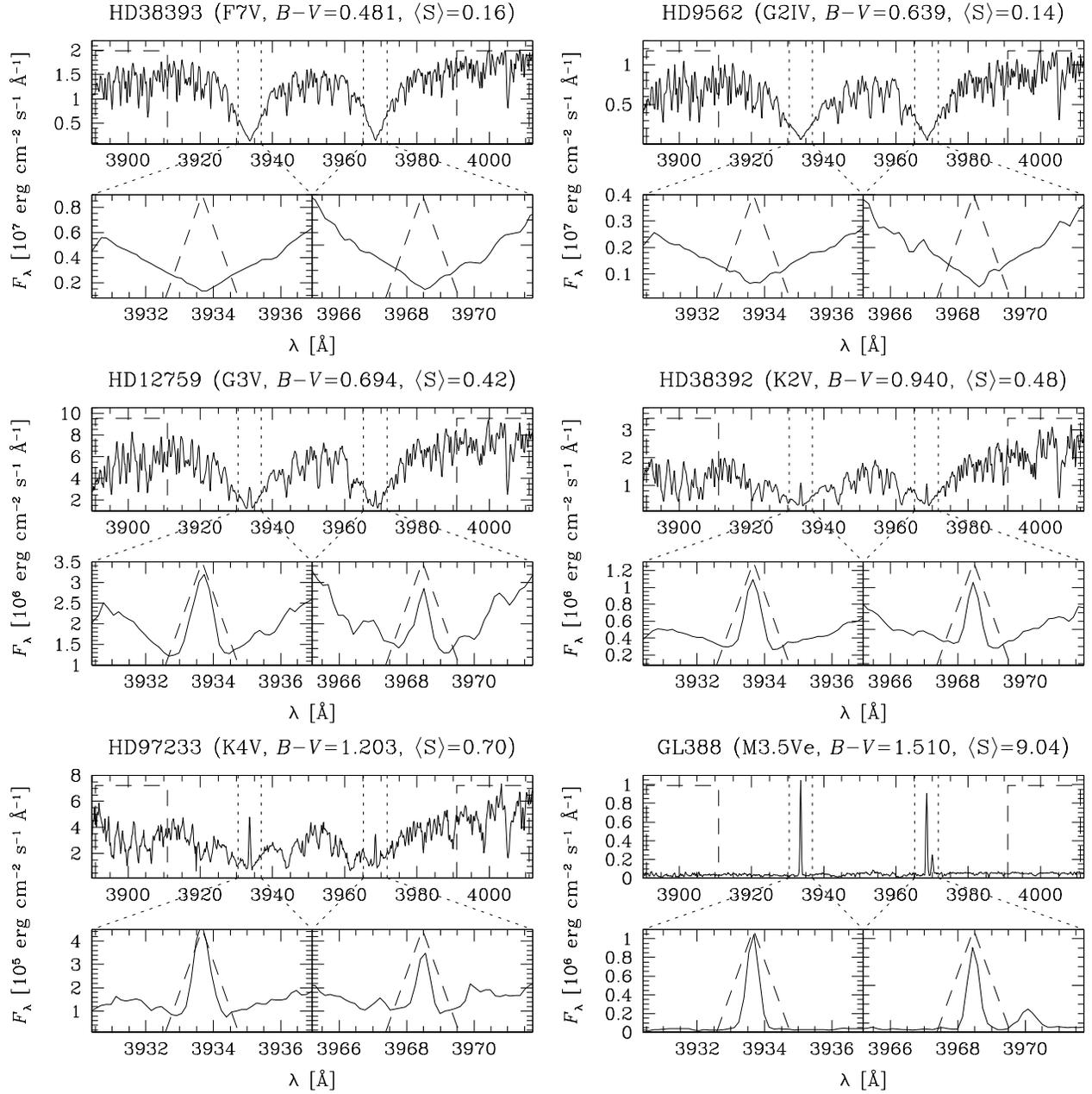} 
\caption{Windows used to compute our $S'$ index, for six stars 
of different spectral types and activity levels. For each star, we show in the 
top panel the whole spectral range used, with the continuum windows indicated 
with dashed lines. Below each of these spectra, we show in detail the H and K 
lines, and the triangular shape used to integrate the fluxes.} 
\label{fig:ventanasS}
\end{figure*}

In Fig.~\ref{fig:ventanasS} we show the windows chosen to integrate the 
fluxes in the \ca\ region, as explained in the next subsections, for several 
stars of different spectral types and different activity levels.

\subsection{Calibration of the \ca\ continuum} \label{sec:contca}

Several authors have constructed empirical stellar fluxes scales, relating 
observed stellar properties, such as colour indexes, to the surface flux in 
specific bandpasses. For example, \citet{1979ApJS...41...47L} found a relation 
between the Johnson $V\!-\!R$ and the surface flux in the 3925--3975~\AA\ 
bandpass. \citet{1988A&A...191..253P} developed an empirical calibration for the 
flux at the \ca\ H and K lines, and \citet{1991A&A...251..199P} did something 
similar for the flux close to the \hal\ line. \citet{1995ApJ...442..778H} 
derived the flux in the \ca\ infrared triplet, H and K lines and \hal\ as a 
function of $b\!-\!y$, and \citet{hall96} have constructed empirical fluxes for 
the continuum near \ca\ H and K and \hal\ as a function of different colour 
indexes. These results apply basically to F-K dwarfs. 

\begin{figure}[t]
\centering 
\includegraphics[width=\columnwidth]{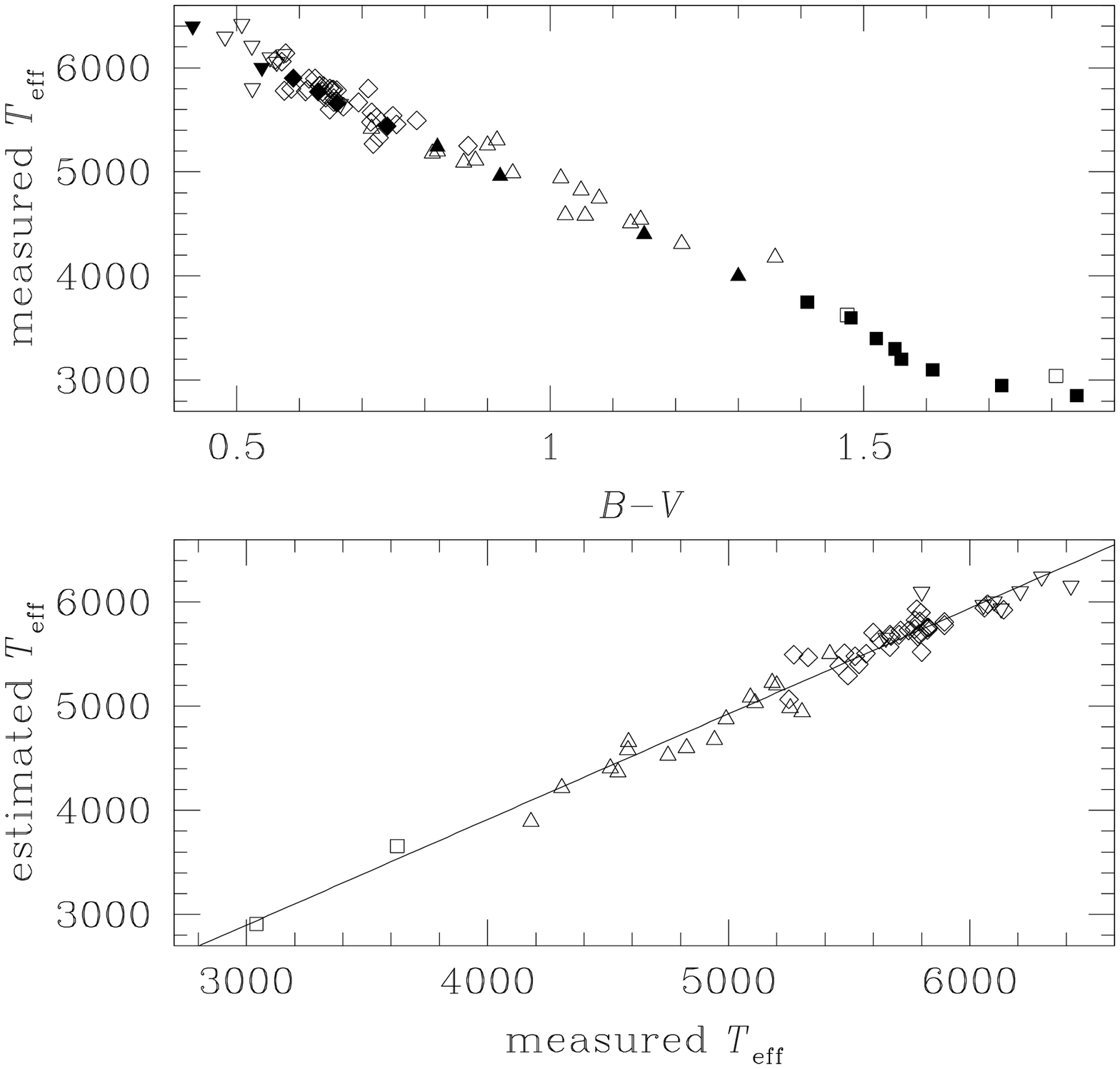} 
\caption{In the upper panel, we show in open symbols the measured \teff\ 
for 62 stars of our sample, taken from \citet{1997A&AS..124..299C, 
2001A&A...373..159C} and \citet{2003A&A...397L...5S}. The filled symbols 
correspond to the data used to interpolate, from \citet{1966ARA&A...4..193J}. 
The lower panel shows the \teff\ calculated from the interpolation as a function 
of the measured \teff , for these 62 stars. The fit is given by 
equation~\ref{eq:Teff}. 
As in the rest of the figures, the symbols $\triangledown$, $\lozenge$, 
$\triangle$ and $\square$ represent F, G, K and M stars respectively. 
}
\label{fig:grafT}\end{figure}

To derive a calibration for the \ca\ continuum directly from our flux-calibrated 
observations, we integrated the continuum flux (as measured from Earth) in two 
passbands near the center of the \ca\ lines, defined in a 
similar way as the continuum passbands of the Mount Wilson Observatory: 
the blue window is centered at 3891~\AA\ an the red one at 4011~\AA , 
and both of them are 20~\AA\ wide. We average this quantities to obtain a mean 
flux at the continuum, $f_{\mathrm{BR}}$, and we converted it to an absolute 
stellar flux, $F_{\mathrm{BR}}$ \citep{2000asqu.book.....C}:
\begin{equation}\label{eq:FBR}
F_{\mathrm{BR}} = \frac{4\pi\sigma\teff^4}{L_\sun} 
\left(10\,\mathrm{pc}\right)^210^{\left(V+BC-M_{\mathrm{bol}}^\sun\right)}  
f_{\mathrm{BR}}  \, ,\end{equation} 
where we used the values given in \citet{1966ARA&A...4..193J} to interpolate 
the effective temperature \teff\ and the bolometric correction $BC$ as functions 
of \col . 62 of the stars of our sample have measurements of \teff\ from 
\citet{1997A&AS..124..299C, 2001A&A...373..159C} and 
\citet{2003A&A...397L...5S}. For these stars, in Fig.~\ref{fig:grafT} we  
compare the measured \teff\ with the one estimated from the interpolation of 
\citeauthor{1966ARA&A...4..193J}'s data. 
The fit shown in the lower panel is given by
\begin{equation}\label{eq:Teff}
\mathrm{estimated}\,\teff=-145.5 +1.01\mathrm{measured}\,\teff\, ,
\end{equation}
with a correlation coefficient $\rho=0.98$. The scatter in the graph is smaller 
than 5\%, and is independient of spectral type.

\begin{figure}
\centering 
\includegraphics[width=\columnwidth]{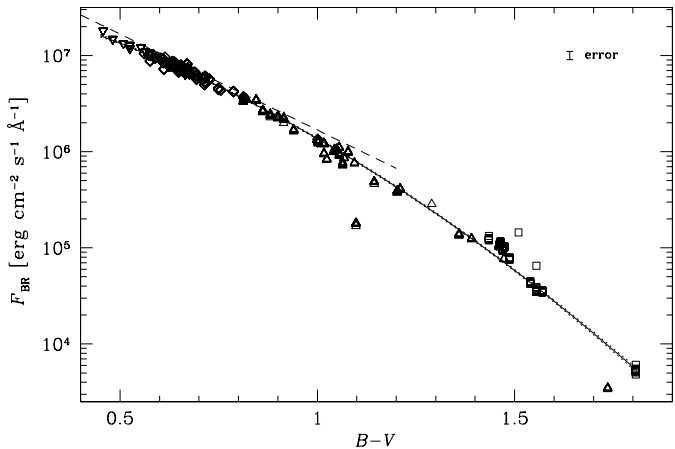} 
\caption{$F_{\mathrm{BR}}$ as a function of colour. The solid line is a 
least-squares fit to the stars excluding HD156425 and HD94683, given by 
equation~\ref{ec:FcontS}. The dashed line is the relation found by 
\citet{hall96}.}
\label{fig:FcontS}
\end{figure}

In Fig.~\ref{fig:FcontS} we plot the absolute mean continuum flux given by 
equation~\ref{eq:FBR} as a function of colour. For the rest of the work, the F 
stars are indicated with triangles pointing downwards, the G stars with 
diamonds, the K stars with triangles pointing upwards and the M stars with 
squares. 

As it can be seen, two K stars clearly apart from the 
general trend: the bluer is HD156425, and the redder HD94683. This star was 
included in our sample because it was classified as a K5V star in the SIMBAD 
database. Nevertheless, according to \citet{1997A&A...323L..49P} its luminosity 
class is III. We can therefore confirm that it is evolved. With regard 
to HD156425, its parallax is unknown, 
so we cannot assure its 
luminosity class. Both stars were discarded in the rest of the analysis.


We did a least-squares fit with a quadratic function, considering errors of 10\% 
in the flux, and obtained:
\begin{equation}\label{ec:FcontS}
\log F_{\mathrm{BR}} = 7.739-0.842 \left( B\!-\!V\right)-0.760 \left( 
B\!-\!V\right)^2 
\end{equation} 
with an error given by
\begin{displaymath}
\sigma _{\left[ \log F_{\mathrm{BR}} \right] } = 
10^{-2} \left[ 2.04 - 8.38 \left( B\!-\!V\right) + 12.59 \left( 
B\!-\!V\right)^2 - \right. 
\end{displaymath}
\begin{equation}
\hspace{18mm}\left. - 8.06 \left( B\!-\!V\right)^3 + 1.88 \left( 
B\!-\!V\right)^4  \right]^{0.5}\, .
\end{equation} 
The function is shown as a full line in Fig.~\ref{fig:FcontS}, and the dotted 
lines represent the $\pm 3 \sigma$ bands. 
This fit has a significance of 99.99\%, considering the 742 spectra with 
$B\!-\!V\le 1.2$. This significance is greatly reduced, however, if all the 886 
spectra, are considered, due to the small number of spectra with $B\!-\!V>1.2$ 
and the increased spread for these points. 

It is possible to introduce a bias in this fit, due to the inhomogeneous 
distribution in colour of our stellar sample, and to the different number of
observations for every star. To investigate this fact, 
we averaged the measurementes in 
bins of 0.1 in colour --excluding HD156425 and HD94683--, to account for the 
bias induced by the larger amount of stars with $B\!-\!V\le1.2$. We computed the 
errors as the standard deviation within each bin, since the bins are narrow 
enough to contain only stars with very 
similar photospheres. Finally, we fitted the results with a quadratic 
function. The differences with Eq.~\ref{ec:FcontS} where always smaller than 
15\%. 

\citet{hall96} also studied the behaviour of the continuum flux at 3950~\AA , 
compiling spectrophotometric data of several authors. He found:
\begin{equation}
\log (F_{3950})_{\mathrm{Hall}} = \left[ -1.995 \, \left( B\!-\!V\right)  + 
8.221 \right] \pm 0.045\, ,
\end{equation} 
for a sample smaller than 95 dwarf stars with $-0.1 \leq B\!-\!V \leq 1.2$. This 
fit is shown as a dashed line in Fig.~\ref{fig:FcontS}. The difference 
between both expressions is probably due to two different facts that have the 
opposite effect.On one hand, the chosen continuum is not the same. At 3950~\AA\  
what is measured is, in fact, a ``pseudo-continuum'': the \ca\ lines are so wide 
that at that wavelength it still does exist some absorption. Our flux, measured 
far away from the cores, is more intense, and certainly it is a more realistic 
estimation of the true continuum, since the contribution of the line absorption 
is smaller. On the other hand, \citet{hall96} corrected the fluxes for the 
blanketing coefficients and we had not, which should result in 
our fluxes being smaller. 

We also performed a linear fit, as was done by \citet{hall96}. 
However, for the stars with $B\!-\!V\le 1.2$ the significance of the linear fit 
is reduced to 95\%. For this reason, we prefer the quadratic form given in Eq. 
2.

\subsection{Calibration of the $S$ index} \label{sec:S}

As far as we know, the only work which compiles measurements of the 
index $S$  for Southern stars is the one from \citet{1996AJ....111..439H}. They 
have observed more than 800~stars with a Cassegrain 
Spectrograph. Although our sample of stars is smaller, we have collected several 
observations for each star, which allows us to study the behaviour in time of a 
particular star \citep[see, for example][]{proxima}. Besides, our sample 
of stars spans a wider range in spectral type, since 
\citeauthor{1996AJ....111..439H} only studied stars with $0.5\le B\!-\!V\le 
1.0$. 

We have defined an $S'$ index in a similar way as the one used in Mount Wilson: 
we integrated the flux in two windows centered at the cores of the \ca\ H and K 
lines, weightened with triangular profiles of 1.09~\AA\ FWHM to mimic the 
response of the Mount Wilson instrument, and computed $S'$ as the ratio between 
this average line flux and the one at the continuum defined in the previous 
section. 

\begin{figure} [b]
\centering
\includegraphics[width=\columnwidth]{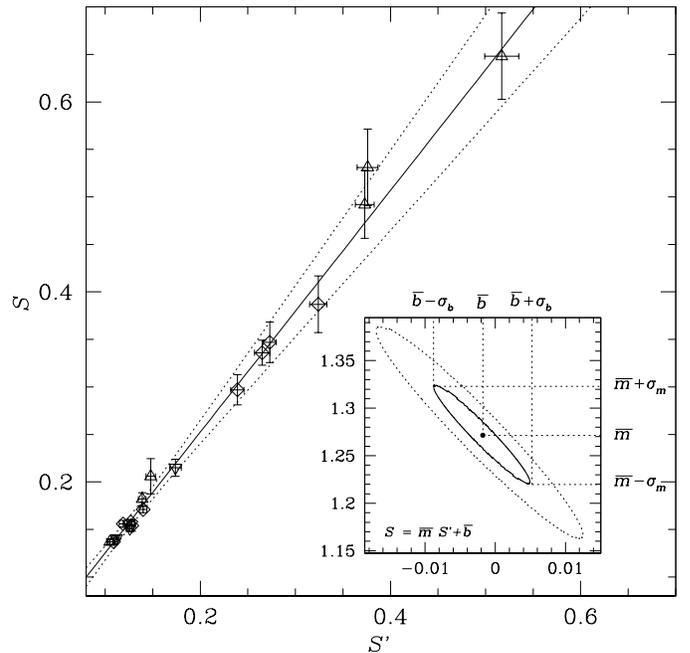}
\caption{$S$ index, as measured in Mount Wilson, as a function of our mean $S'$ 
index, for the 18 stars chosen as reference stars. The full line is a 
least-squares fit, and the dotted lines are the $\pm3\sigma$ curves. In the 
insert we indicate the 39.3\% confident level (full line) and the 90\% 
confidence level (dotted line) for the two parameters fitted.}
\label{fig:SvsSp}
\end{figure}

In order to corroborate the accuracy of this definition, we have 
specially included in our sample 18 stars from the Mount Wilson database defined 
as non-variable by \citet{1996AJ....111..439H}, and we have observed each 
of these stars between 8 and 14 times. 
In Fig.~\ref{fig:SvsSp} we plot the Mount Wilson $S$ index as a function of our 
mean $S'$ for these stars, calculated by weigthing each individual measurement 
with an error of 10\%. The data used for this figure are presented in 
Table~\ref{tab:estandares}. We made a least-squares fit 
obtaining:
\begin{equation} \label{ec:SvsSp}
S = \left( 1.271 \pm 0.052 \right) 
S'- \left( 0.002  \pm 0.007 \right) \, ,
\end{equation} 
with a correlation coefficient of -0.96. The significance of the fit is larger 
than 87\%. As can be seen in Table~\ref{tab:estandares}, the color 
range involved in this calibration goes from \col = 0.48 to \col = 0.94, only a 
portion of the whole color range of our sample, 0.45$\leq$\col $\leq$1.81. 
Therefore, this technique allows us to reproduce the Mount Wilson measurements 
with our instrument, with great confidence for stars with 
\col$<$ 1, and with some uncertainty for the whole sample. Since up to 
present there are no systematic measurements of M stars, we cannot estimate the 
errors introduced by this procedure. 

\begin{table}
\caption{Stars used to calibrate our $S'$ index with the Mount Wilson 
$S$, as shown in Fig.~\ref{fig:SvsSp}. $N$ is the number of spectra available 
for each star. }  \label{tab:estandares}\centering
\begin{tabular}{lcccrcc} 
\hline\hline 
star & \col\ & $S$ & $\sigma_S$ & $N$ & $S'$ & $\sigma_{S'}$ \\ \hline
HD38393  & 0.481 & 0.151 & 0.003 & 13 & 0.126 & 0.004 \\
HD16673  & 0.524 & 0.215 & 0.009 & 10 & 0.174 & 0.006 \\
HD45067  & 0.564 & 0.141 & 0.002 & 11 & 0.112 & 0.003 \\
HD30495  & 0.632 & 0.297 & 0.016 & 12 & 0.239 & 0.007 \\
HD9562   & 0.639 & 0.137 & 0.003 &  9 & 0.109 & 0.004 \\ 
HD11131  & 0.654 & 0.336 & 0.013 & 10 & 0.265 & 0.008 \\
HD1835   & 0.659 & 0.347 & 0.021 & 14 & 0.273 & 0.007 \\
HD158614 & 0.715 & 0.158 & 0.003 & 12 & 0.127 & 0.004 \\
HD3443   & 0.715 & 0.182 & 0.007 & 10 & 0.139 & 0.004 \\
HD3795   & 0.718 & 0.156 & 0.004 & 10 & 0.119 & 0.004 \\
HD10700  & 0.727 & 0.171 & 0.003 & 12 & 0.140 & 0.004 \\
HD152391 & 0.749 & 0.387 & 0.030 & 12 & 0.324 & 0.009 \\
HD219834 & 0.787 & 0.154 & 0.006 & 12 & 0.128 & 0.004 \\
HD26965  & 0.820 & 0.206 & 0.019 & 10 & 0.148 & 0.005 \\
HD17925  & 0.862 & 0.648 & 0.046 &  8 & 0.517 & 0.018 \\
HD22049  & 0.881 & 0.492 & 0.036 & 14 & 0.373 & 0.010 \\
HD23249  & 0.915 & 0.137 & 0.003 & 14 & 0.105 & 0.003 \\
HD38392  & 0.940 & 0.531 & 0.040 & 12 & 0.376 & 0.011 \\ \hline

\hline 
\end{tabular}
\end{table}

\subsection{Conversion factor between $S$ and the flux}

The Mount Wilson $S$ index can be converted to the average surface flux in the 
\ca\ lines through the relation: 
\begin{equation}
F_{\mathrm{HK}} = F_{\mathrm{bol}} 1.34\, 10^{-4}\, S\, C_{\mathrm{cf}} 
\, ,
\end{equation} 
where $C_{\mathrm{cf}} \left( B\!-\!V\right)$ is a conversion factor which 
depends on colour. Two different expressions are widely used for this 
factor, the first one given by \citet{1982A&A...107...31M} and corrected by 
\citet{1984ApJ...279..763N} and the other one given by 
\citet{1984A&A...130..353R}. 
The deductions used in both works to derive $C_{\mathrm{cf}}$ involve complex 
calibration procedures. 

The conversion factor computed by \citeauthor{1982A&A...107...31M} and 
\citeauthor{1984ApJ...279..763N} is valid for $0.45\leq B\!-\!V \leq 1.5$, and is given by:
\begin{equation} \label{ec:cfMN}
\log\, C_{\mathrm{cf}}^{\mathrm{MN}} = 1.13\, y^3 - 3.91\, 
y^2 + 2.84\, y -0.47 + \Delta 
\end{equation} 
where $y\equiv (B\!-\!V)$ and the last term is, as a function of $x\equiv 
0.63-(B\!-\!V)$:
\begin{equation}
\Delta\, \left(x\right)= \left\lbrace \begin{array}{ll} 0 & 
\mathrm{if}\; x<0 \\ 0.135\, x - 0.814\, x^2 + 6.03\, x^3 & \mathrm{if}\; 
x>0\, .\end{array} \right. 
\end{equation} 
On the other hand, the 
conversion factor computed by \citeauthor{1984A&A...130..353R} is:
\begin{equation} \label{ec:cfR}
\log\, C_{\mathrm{cf}}^{\mathrm{R}} = 0.25\, y^3 - 1.33\, 
y^2 + 0.43\, y +0.24
\end{equation} 
and it is valid for main-sequence stars with $0.3\leq B\!-\!V \leq 1.6$. These 
conversion factors have also been used in recent works 
\citep[see, for example, ][]{2004ApJS..152..261W}. 

Since we have simultaneous measurementes of the $S$ index and the core 
fluxes, we can calculate directly the correction factor as a function of the 
index and the flux, fitting the expression:
\begin{equation}\label{ec:factor_correccion}
\log C_{\mathrm{cf}} = \log F_{\mathrm{HK}} - \log \sigma \teff ^4 -\log S - 
\log 1.34 + 4\, ,
\end{equation} 
were we have obtained $S$ as a function of the measured $S'$ using 
Eq.~\ref{ec:SvsSp}. 

\begin{figure}\centering
\includegraphics[width=\columnwidth]{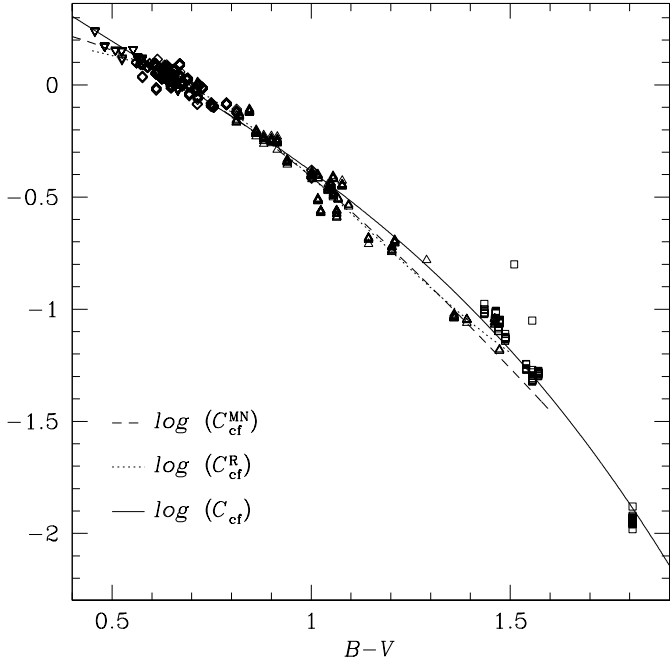}
\caption{Conversion factor between Mount WIlson $S$ index and $F_{\mathrm{HK}}$. 
The dashed line corresponds to \citet{1984ApJ...279..763N}, the dotted line to 
\citet{1984A&A...130..353R}, and the full line to our derived factor, 
Eq.~\ref{ec:factorcorr}. The stars HD156425 and HD94683 have been discarded. }
\label{fig:factor_corr}
\end{figure}

In Fig.~\ref{fig:factor_corr} we plot the correction factor computed from 
Eq.~\ref{ec:factor_correccion} as a function of colour. The full line 
corresponds to a least-squares fit to the data, with a significance higher  
than 99\% for the 880 spectra included. This fit is given by:
\begin{equation}\label{ec:factorcorr}
\log C_{\mathrm{cf}} = -0.33 y^3 + 0.55 y^2 - 1.41 y + 0.8\, ,
\end{equation}
where $y\equiv B\!-\!V$, and the errors of the fit are:
\begin{displaymath}
\sigma_{\left[ \log C_{\mathrm{cf}}\right] } = \left[ 0.24 -1.48 y 
+ 3.65 y^2  -4.7 y^3 + 3.34 y^4 -1.24 y^5 +\right. 
\end{displaymath} \vspace*{-4mm}
\begin{equation}
\hspace{15mm} \left.   + 0.19 y^6 \right] ^{1/2} 10^{-1} \, .
\end{equation}
This fit is valid for the whole range $0.45\le $ \col\ $\le 1.81$. 
Also shown in the same figure are the conversion factors of 
\citet{1982A&A...107...31M} and \citet[][dashed line]{1984ApJ...279..763N}
and \citet[][dotted line]{1984A&A...130..353R}. The three conversion factors 
are very similar for stars with $0.6\le $ \col\ $\le 1.0$, which are the most 
crowded of our sample. Our factor slightly aparts from the others for $1.0\le $ 
\col\ $\le 1.5$, while it extends the calibration up to \col\ $=1.8$. 

It is worth noting that the conversion factor of Eq.~\ref{ec:factorcorr} 
was derived using spectra which were calibrated in flux, the same which were  
used to measure the $S$ index. On the other hand, 
the ones given by \citet{1982A&A...107...31M} and \citet{1984ApJ...279..763N}, 
and \citet{1984A&A...130..353R}, were computed using interpolated fluxes 
at the appropiate wavelengths from other instruments.


\subsection{The photospheric contribution}

The $S$ index depends both on the chromospheric and photospheric radiation, 
while the activity is related only to the chromospheric part. This dependence of 
$S$ is due, in the first place, to the flux in the continuum windows, which 
is a strong function of the spectral type. Furthermore, the passbands centered 
in the cores of the lines include some photospheric contribution from the flux, 
in particular outside the H$_1$ and K$_1$ minima. For these reasons it 
is usually considered that a better indicator of chromospheric activity is 
$R_{\mathrm{HK}}'$, defined as:
\begin{equation}
R_{\mathrm{HK}}' = \frac{F_{\mathrm{HK}}-F_{\mathrm{phot}}}{F_{\mathrm{bol}}}\, 
,
\end{equation} 
where
$F_{\mathrm{HK}}$ is the average surface flux in the center of the H and K 
lines, $F_{\mathrm{phot}}$ is the photospheric contribution integrated in the 
lines and $F_{\mathrm{bol}} = \sigma T_{\mathrm{eff}}^4$. 

The triangular profiles used to integrate the line fluxes $F_{\mathrm{HK}}$ 
include some flux outside the H$_1$ an K$_1$ minima, which is formed in the 
stellar atmosphere below the temperature minimum. This flux of photospheric 
origin, $F_{\mathrm{phot}}$, introduces a dependence on stellar colour which
has to be substracted in order to obtain the net chromospheric line flux 
$F'_{\mathrm{HK}}$ defined as:
\begin{equation}
F'_{\mathrm{HK}} = F_{\mathrm{HK}} - F_{\mathrm{phot}}\, .
\end{equation} 

Unfortunately, our instrumental resolution is not good enough to measure
the photospheric flux in this way.

We have adopted instead the functional form proposed by 
\citet{1984ApJ...279..763N}:
\begin{equation}\label{ec:smin}
\log \left( \frac{F_{\mathrm{phot}}}{F_{\mathrm{bol}}} \right) = -4.02 -1.40 
\left( B\!-\!V \right) \, ,
\end{equation} 
in the range $0.44 < B\!-\!V < 0.82$. This expression has been derived for the 
Mount Wilson survey. As the authors explain in the cited work, they have also 
used the same form for \col\ $>0.82$. In fact, it becomes negligible for \col\ 
$\gtrsim 1.0$, so we extrapolated it to our full range of colours. We will use 
this expression to obtain the chromospheric flux used for the analysis of 
section~\ref{sec:AyS}.


\section{\hal\ index}\label{sec:Hal}

Although the \ca\ resonance lines have been the most widely used chromospheric 
indicators, they present two problems when they are used to study cool stars. 
First, the coolest stars --mid K to mid M-- are progressively redder than the 
hottest ones, and therefore the spectral range where these lines are located becomes 
less intense --relatively to other ranges of the visual spectrum-- as the 
temperature decreases. 
Moreover, these coolest stars are instrinsically faint, and therefore this 
problem is increased. As a consequence, the \ca\ lines are not the most adequate 
to observationally study the coolest stars of our sample, because the 
signal-to-noise ratio for these lines is very small, even whith long exposure 
times. 


Although in most solar-type dwarfs the formation of the \hal\ line is 
dominated by photoionization, as we move to later stars 
the decrease in the photospheric radiation temperatures gives place to
a significant collisional contribution to the source function 
and eventually \hal\ goes into emission for the dKe and 
dMe stars \citep{1993MNRAS.262....1T,1994A&A...281..129M,1997A&A...326..249M}. 
Therefore, \hal\ is also usually used as a chromospheric indicator 
\citep{1986ApJS...62..241L, 1991A&A...251..199P, 1993MNRAS.262....1T, 
1995A&A...294..165M,  1995ApJ...442..778H}. Furthermore, the spectral location 
of this line is much more adequate for late stars and the integration times 
needed to obtain a proper signal-to-noise ratio are much lower than for the \ca\ 
lines. 

\begin{figure*}
\centering 
\includegraphics[width=17cm]{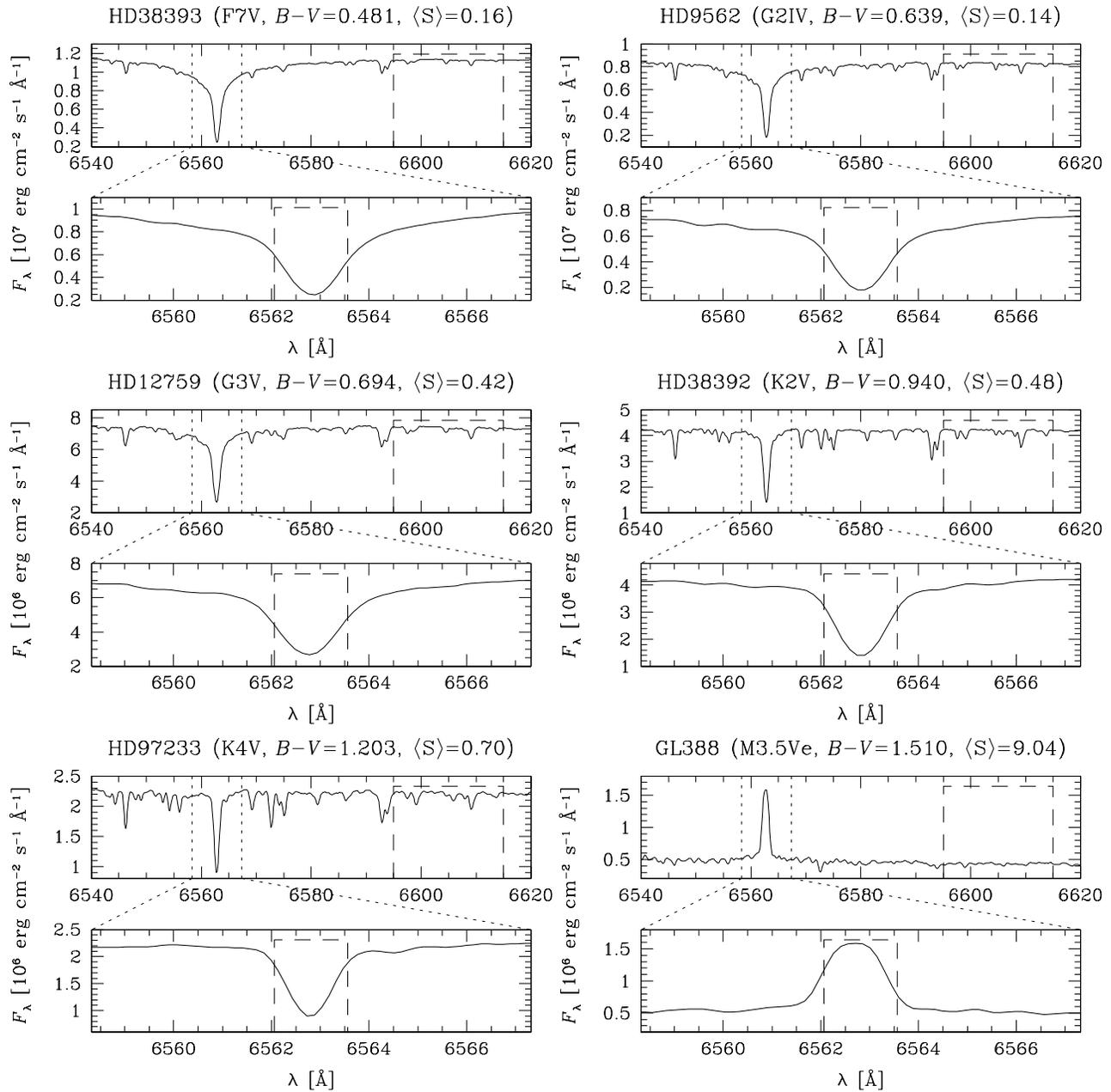} 
\caption{Windows used to measure the fluxes in \hal , for six 
stars of different spectral types and activity levels, the same as in 
Fig.~\ref{fig:ventanasS}. For each star, we show in the top panel the whole 
range used, with the continuum window indicated with dashed lines. Below each of 
these spectra, we show in detail the \hal\ line, superimposed with the 
filter used to integrate the fluxes.} \label{fig:ventanasA}
\end{figure*}

In Fig.~\ref{fig:ventanasA} we show the windows of interest for \hal\ 
for the same stars than in Fig.~\ref{fig:ventanasS}. In the next 
subsections we explain how we used these observations. 

\subsection{Calibration of the \hal\ continuum}\label{sec:FcontHal}

In a similar way to the calcium index, we choose two windows to integrate the 
fluxes, one centered in the line and the other in the continuum nearby. We  
calibrated the continuum for this line as we did in Section~\ref{sec:contca}, 
on one hand, to corroborate the quality of the flux-calibration in 
the \hal\ region and, on the other, because an expression of this kind can be 
used to calibrate in flux normalized spectra in this spectral region. 

\begin{figure}[b]
\centering
\includegraphics[width=\columnwidth]{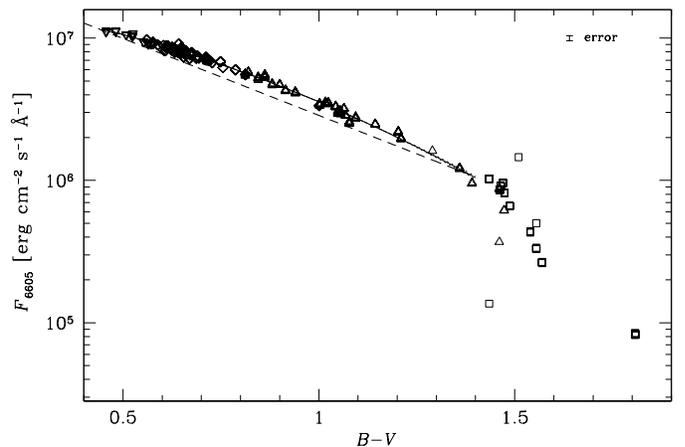}
\caption{$F_{6605}$ as a function of colour. The dashed line is the 
relation found by \citet{hall96}. 
}
\label{fig:FcontA}
\end{figure}

We measured the continuum flux averaged in a window 20~\AA\ wide 
centered at 6605~\AA , and we translated it to surface flux, using an 
equation similar to Eq.~\ref{eq:FBR}. It is plotted as a function of colour 
in Fig.~\ref{fig:FcontA}. 
\col\ is not the best \teff\ proxy for the coolest stars, as can
be seen from the notorious scatter for the later stars of 
Fig.~\ref{fig:FcontA}, but it is the only color index available for our whole 
sample ofstars. Since this calibration can induce errors in the \hal\ flux, we 
restrict our analysis to stars with \col $\leq 1.4$, and we indicate 
the stars excluded from the analysis with a note in Table~\ref{tab:stars}. 
For the 783 points corresponding to these stars 
we found: 
\begin{equation}
\log F_{6605}= 7.281 - 0.298 \left( B\!-\!V \right) -0.431 \left( B\!-\!V 
\right)^2\, ,
\end{equation} 
with errors:
\begin{displaymath}
\sigma_{\left[ \log F_{6605} \right] } =10^{-2} \left[ 1.06 - 5.08 \left( 
B\!-\!V \right) + 8.97 \left( B\!-\!V \right)^2 - \right.
\end{displaymath}
\begin{equation}
\hspace{17mm}\left. - 6.96 \left( B\!-\!V \right)^3 
+ 2.00\left( B\!-\!V \right)^2 \right]^{0.5}
\end{equation}
and a significance of 84\%, assuming individual errors of 4\%. 
This fit is drawn with a full line in the figure. \citet{hall96} compiled data 
from different sources of spectrophotometry, and, using a method that 
interpolates the continuum flux to the wavelength of the line center, found a 
similar behaviour:
\begin{equation} 
\log\, \left( F_{6563} \right)_{\mathrm{Hall}} = 
\left [ -1.081 (B-V) +7.538 \right] \pm 0.033\; \mathrm{dex}\, ,
\end{equation}
which is marked with a dashed line in the same figure. 



\subsection{\hal\ flux and its photospheric contribution} \label{sec:FHa}

\begin{figure}
\centering
\includegraphics[width=\columnwidth]{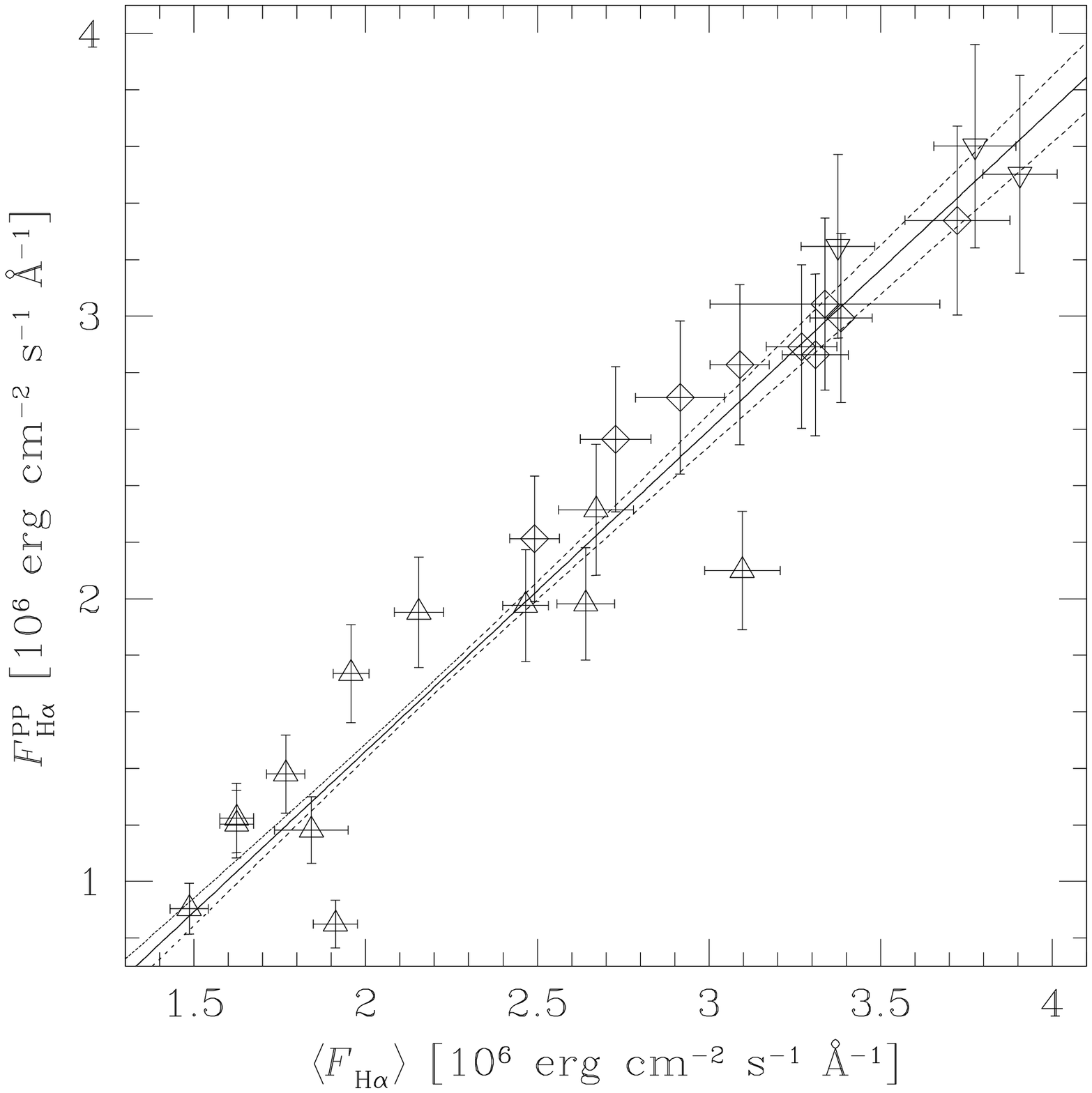}
\caption{Average surface flux in the center of \hal , for the 24 stars in common 
between our sample ($\langle F_{\mathrm{H}\alpha} \rangle$) and the one of 
\citet[$F^{\mathrm{PP}}_{\mathrm{H}\alpha}$]{1991A&A...251..199P}. The full 
line corresponds to a least-squares fit of the data, and the 
dashed line to the $\pm 3\sigma$ band. }
\label{fig:FHAPP}
\end{figure}

We computed the flux in the \hal\ line, $F_{\mathrm{H}\alpha}$, as the average 
surface flux in a 1.5~\AA\ square passband centered in the line.
\citet{1991A&A...251..199P} studied the flux on the stellar surface in a 
1.7~\AA\ window centered at \hal , for a sample of G and K dwarf stars, 24 of 
which were also included in our sample. For these stars, we calculated $\langle 
F_{\mathrm{H}\alpha} \rangle$, the average flux in the line, which are shown in 
Fig.~\ref{fig:FHAPP}. By means of a least-squares fit, we find that the 
\citeauthor{1991A&A...251..199P} fluxes $F^{\mathrm{PP}}_{\mathrm{H}\alpha}$ are 
related to ours $\langle F_{\mathrm{H}\alpha} \rangle$ by:
\begin{equation} 
F^{\mathrm{PP}}_{\mathrm{H}\alpha}=(1.136 \pm 
0.063)\, \langle F_{\mathrm{H}\alpha} \rangle - (0.811 \mp 0.139)\, ,
\end{equation} 
where \citeauthor{1991A&A...251..199P} values have a 10\% errors, and we have 
considered the \emph{individual} errors of our measurements to be 10\% as well. 
However, since we have 
between 5 and 14 observations for most stars, the errors in Fig.~\ref{fig:FHAPP} are much 
smaller. 

As it can be seen in Fig.~\ref{fig:FHAPP}, the correlation between both 
sets of measurements is excelent, particularly considering that our fluxes and the 
\citeauthor{1991A&A...251..199P} ones were taken at different moments, i.e. with
different levels of activity.

To estimate the photospheric contribution to this line is much more complicated 
than for the calcium ones. On one hand, for the Ca lines one can determine where 
in the wings the flux is of photospheric origin. On the other, the photospheric 
contribution to the \ca\ line fluxes is less important than the chromospheric 
one, while the center of the \hal\ line is clearly not opaque even for the least 
active stars, and therefore in this case the chromospheric contribution is the 
result of a subtraction between two comparable quantities. 

\begin{figure}
\centering
\includegraphics[width=\columnwidth]{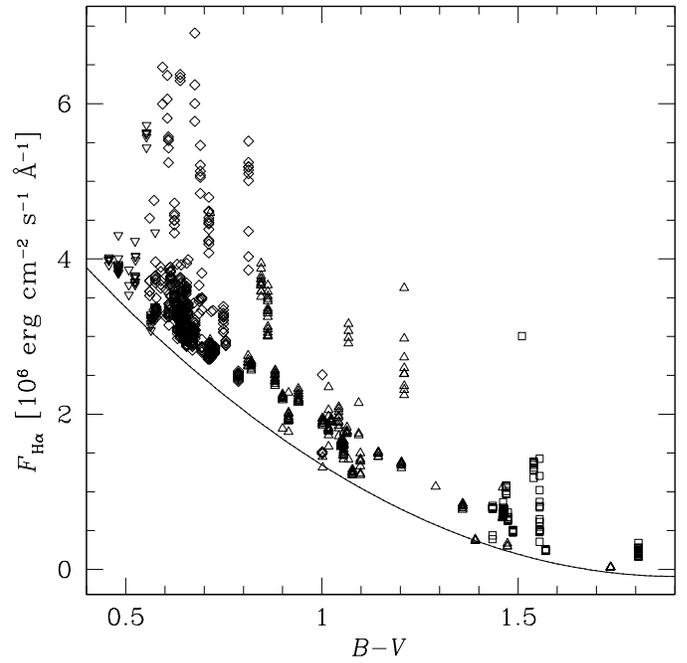}
\caption{Average surface flux in the center of \hal\ as a function of  
colour, for the whole sample of stars. The full line represents 
the minimum flux given by Eq.~\ref{ec:Amin}. }
\label{fig:Amin}
\end{figure}

An usual way of overcoming this problem, when the number of observations is 
large enough, is to assume that the photospheric flux depends only on 
stellar colour and corresponds to the minimum flux observed. In 
Fig.~\ref{fig:Amin} we show $F_{\mathrm{H}\alpha}$ as a function of \col . The 
full line is the curve that represents the minimum flux of the line, which we 
fit with a quadratic polynomial of the form:
\begin{equation} \label{ec:Amin}
F_{\mathrm{H}\alpha}^{\mathrm{min}} = 10^4 \left[ 629-670 \left( 
B\!-\!V \right) +176 \left( B\!-\!V \right)^2 \right] \, .
\end{equation}  

\subsection{Relation between the \ca\ and \hal\ fluxes}
\label{sec:AyS}

It is usually accepted that there is a tight relation between the 
chromospheric fluxes emitted in \hal\ and in the H and K \ca\ lines. However, most works 
where this relation has been observed \citep[see, for example,] [] 
{1989ApJ...345..536G, 
1990ApJS...74..891R, 1990ApJS...72..191S, 1995A&A...294..165M} found it by using 
averaged fluxes for both the calcium and the hydrogen lines, which were not 
obtained simultaneously, and were even collected from different sources. 
\citet{1993MNRAS.262....1T}, on the other hand, did use simultaneous 
observations, which represent a particular moment of each star and not an 
average behaviour, but they observed each star only once. 

\begin{figure}[t]
\centering
\includegraphics[width=\columnwidth]{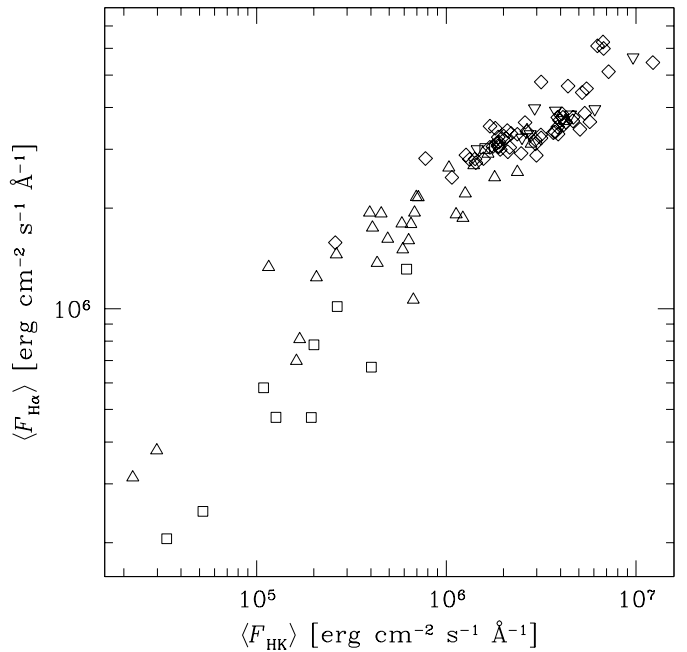}
\caption{Average \hal\ surface fluxes as a function of the average \ca\ surface 
fluxes. 
}
\label{fig:AyStot}
\end{figure}

In Fig.~\ref{fig:AyStot} we have reproduced this kind of studies averaging 
for each star our measurements of the \ca\ and the \hal\ fluxes, 
obtained by direct integration of the spectra, and weighting them with 
their individual errors. 
For the 108 points shown there is, in fact, a quite clear correlation 
between both fluxes, specially for the stars whith the strongest chromospheric 
emission. 

\begin{figure*}[t]
\centering
\includegraphics[width=17cm]{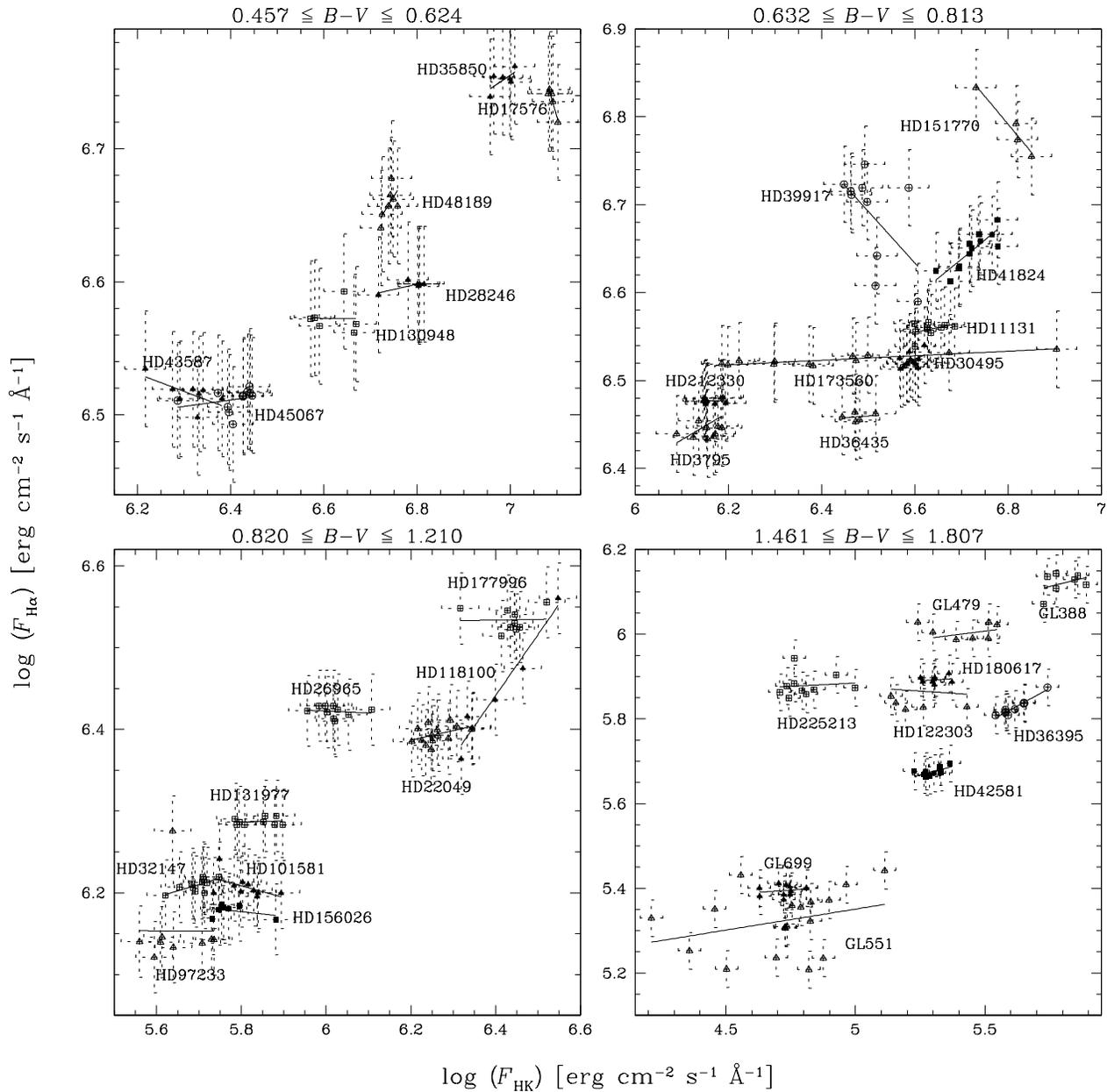}
\caption{\hal\ surface fluxes as a function of the \ca\ 
surface fluxes, for stars of different spectral types, divided into different 
color bins, as indicated. }\label{fig:bins}
\end{figure*}

However, the situation is different when the individual 
stars are studied separately. In Fig.~\ref{fig:bins} we plot the 
individual simultaneous measurementes of each flux for several stars of 
different spectral types and different levels of activity, divided into color 
bins as stated. The error bars represent the 10\% of uncertainty we asign to our 
flux calibration, as explained in \citet{2004A&A...414..699C}. We also show 
the linear fits for each star. It can be seen that the behavior is different in 
each case: in some stars both fluxes are well correlated, as in HD118100 and 
HD36395, although the slopes of the fits are not the same. In other stars as 
HD30495 $F_{\mathrm{H}\alpha}$ seems to be almost independent of the level of 
activity measured in the \ca\ lines, and there are even stars like HD151770 
where the fluxes are anti-correlated. 

We found no evidence of dependence of this behaviour with spectral type 
or level of activity. In fact, even when we restrict the analysis to dMe stars, 
where the mechanisms of formation of the lines are supposed to be similar, the 
behaviour differs for different stars: for example, for GL551 (Prox Cen) we 
found a good correlation for both fluxes \citep[see][where 
the study have been made much more carefully taking into account 
variations produced by flares]{proxima} while for GL699 (Barnard's star) \hal\ 
seems to be independent of the activity measured by \ca . In 
Table~\ref{tab:stars} we included in the last two columns the slopes and 
correlation coefficients of these fits. It can be seen that there is no 
clear tendency of how \hal\ varies with \ca . 


This fact seems to indicate that the trend present in Fig.~\ref{fig:AyStot} is 
not the product of a direct relationship between the activity measured in \hal\ 
and in the \ca\ lines, but it is due to a correlation 
of each separate flux with the stellar colour or the spectral type.

\begin{figure}[t]
\centering
\includegraphics[width=\columnwidth]{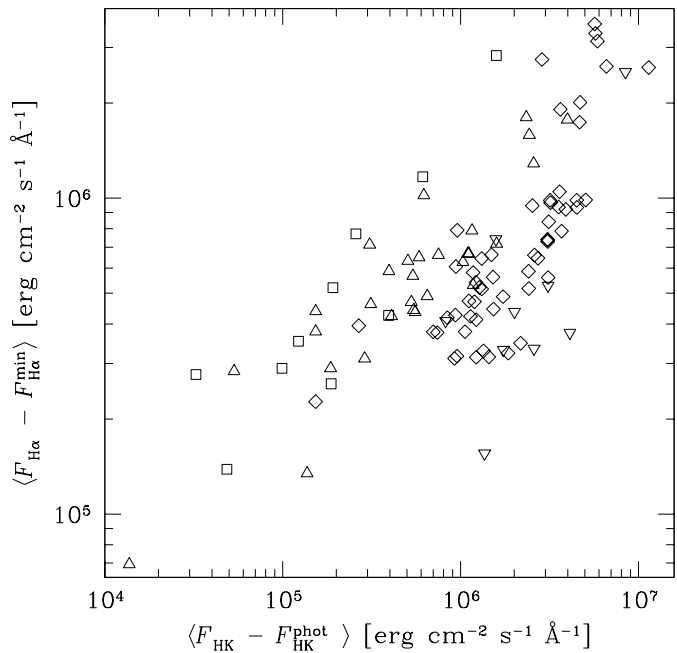}
\caption{Average surface fluxes of chromospheric origin, for \hal\ and the H 
and K lines. The tight relation of Fig.~\ref{fig:AyStot} has a much larger 
spread. 
}\label{fig:AySmin}
\end{figure}

Finally, 
in Fig.~\ref{fig:AySmin} we plot the fluxes directly related to the 
chromospheric activity, $\langle F_{\mathrm{HK}} - 
F_{\mathrm{HK}}^{\mathrm{phot}} \rangle$ and $\langle F_{\mathrm{H}\alpha} - 
F_{\mathrm{H}\alpha}^{\mathrm{min}} \rangle$, where we used Eq.~\ref{ec:smin} 
and \ref{ec:Amin} to compute $F_{\mathrm{HK}}^{\mathrm{phot}}$ and 
$F_{\mathrm{H}\alpha}^{\mathrm{min}}$ respectively. We see that the correlation 
becomes much more unclear and the spread much larger. 
\citet{1993ApJS...85..315S} have already argued that, given the different 
conditions of formation of \hal\ and the H and K lines, it is possible to find 
significant differences between their behaviour when observing two different 
stars, since the underlying physics for each line could differ much between 
them.

\section{Summary}\label{sec:sumario}

This work concentrates on the statistical study of two activity indicators for 
stars of different activity levels and spectral types. We use mid-resolution 
spectra calibrated in flux for 109 Southern stars from F6 to M5. We have a total 
of 917 simultaneous observations of these stars, which include from \ca\ H and K 
to the \hal\ region.

The first indicator we have characterized is the one related to the H and K \ca\ 
lines. We studied the dependence of the continuum flux near these lines with colour, 
and we found an excellent agreement with similar studies by other authors, 
corroborating the quality of our observations and reduction techniques. Using 18 
stars selected because of their small variability, our $S'$ index correlates 
very well with the index from the Mount Wilson Observatory, which allows us to 
intercalibrate both sets of measurements. 

Usually, the $S$ index is measured with the Mount Wilson instrument or from 
normalized spectra, and a conversion factor is computed --through indirect 
calibrations-- to translate the index to the flux in the line. To our knowledge, 
this is the first work where the $S$ index is measured from flux-calibrated 
spectra, a fact that allows us to have simultaneous observations of the index 
and of the flux in the lines. Taking adventage of this fact, we derived the 
conversion factor between $S$ and the line flux directly, and extended it to a 
wider colour range. 


We also studied the behaviour of the flux in the \hal\ region.  We first 
explored the relation of the continuum flux with colour reproducing very well 
previous results of the literature. In this case we also found an excellent 
match between our flux in the line and the fluxes published by other authors, 
for 24 stars of the sample. 

Usually the \hal\ and \ca\ fluxes are used interchangeably as 
activity indicators. Since we have a large number of simultaneous 
observations of both fluxes we could explore their correlation in detail. 
We found a strong correlation between the mean fluxes of each line for each 
star.
However, when we investigate this relation for 
individual observations of particular stars, the general trend is lost and each 
star shows a particular behaviour, ranging from tight correlations with 
different slopes to anti-correlations, including cases where no correlations are 
found. We found no evidence of dependence of this behaviour with spectral type 
or level of activity. 

We conclude that the tight relationship that is present 
for the mean values is basically the product of the dependence of the mean fluxes with 
stellar colours. Therefore, we found that in general the activity measured in 
\hal\ is not equivalent to the one measured in the \ca\ lines for the whole 
sample of stars. 


 
\begin{acknowledgements}
The CCD and data acquisition system at CASLEO has been
partly financed byR. M. Rich through U.S. NSF grant AST-90-15827. This work made 
 extensive use of the SIMBAD database, operated at CDS, Strasbourg, France. We 
thank an anonymous referee for his/her useful comments.
\end{acknowledgements}

\bibliographystyle{aa}
\bibliography{indices}
 
 

\clearpage
\onecolumn

\begin{center}
\tablefirsthead{
\hline \hline 
\multicolumn{1}{|c}{HD} & \multicolumn{1}{l}{GJ/GL} & Name  &       
& Spectral & \multicolumn{2}{c}{\hspace{-2mm} $V$} &    
\multicolumn{3}{c}{\hspace{-4mm} \col\ } &      
\multicolumn{3}{c}{\hspace{-2mm} Parall.} & 
\multicolumn{2}{c}{\hspace{-4mm} \gravity\ } &      
\multicolumn{3}{c}{\hspace{-4mm} \metal\ } &   
\multicolumn{4}{c}{\hspace{-4mm} \prot\ } & $N$ & 
\multicolumn{2}{c}{$\langle S \rangle$} & 
\multicolumn{2}{c}{$\langle F_{\mathrm{HK}} \rangle$} &  
\multicolumn{2}{c}{$\langle F_{\mathrm{H}\alpha} \rangle$} &
\multicolumn{2}{c}{$m$} &
\multicolumn{2}{c|}{$\rho$} \\ 
& & & & type & \multicolumn{2}{c}{ } & \multicolumn{2}{c}{ } & & 
\multicolumn{2}{c}{ } & & \multicolumn{2}{c}{ } & \multicolumn{2}{c}{ } & & & 
\multicolumn{2}{c}{ } & & & \multicolumn{2}{c}{ } & \multicolumn{2}{c}{10$^{4}$} 
& \multicolumn{2}{c}{10$^5$} &
\multicolumn{4}{c|}{} \\  
\multicolumn{25}{|c}{} & \multicolumn{4}{c}{erg 
cm$^{-2}$ s$^{-1}$ \AA $^{-1}$} &
\multicolumn{4}{c|}{} \\ 
\hline}
\tablehead{
\hline \hline  
\multicolumn{33}{|r|}{\emph{Tab.~\ref{tab:stars} continued}} \\  
\multicolumn{1}{|c}{HD} & \multicolumn{1}{l}{GJ/GL} & Name  &       & Spectral & 
\multicolumn{2}{c}{\hspace{-2mm} $V$} &    \multicolumn{3}{c}{\hspace{-4mm} 
\col\ } &      \multicolumn{3}{c}{\hspace{-2mm} Parall.} & 
\multicolumn{2}{c}{\hspace{-4mm} \gravity\ } &      
\multicolumn{3}{c}{\hspace{-4mm} \metal\ } &   
\multicolumn{4}{c}{\hspace{-4mm} \prot\ } & $N$ & 
\multicolumn{2}{c}{$\langle S \rangle$} & 
\multicolumn{2}{c}{$\langle F_{\mathrm{HK}} \rangle$} &  
\multicolumn{2}{c}{$\langle F_{\mathrm{H}\alpha} \rangle$} &
\multicolumn{2}{c}{$m$} &
\multicolumn{2}{c|}{$\rho$} 
\\ 
& & & & type & \multicolumn{2}{c}{ } & \multicolumn{2}{c}{ } & & 
\multicolumn{2}{c}{ } & & \multicolumn{2}{c}{ } & \multicolumn{2}{c}{ } & & & 
\multicolumn{2}{c}{ } & & & \multicolumn{2}{c}{ } & 
\multicolumn{2}{c}{10$^4$ } & 
\multicolumn{2}{c}{10$^5$ } &
\multicolumn{4}{c|}{}
\\ \multicolumn{25}{|c}{} & \multicolumn{4}{c}{erg 
cm$^{-2}$ s$^{-1}$ \AA $^{-1}$} &
\multicolumn{4}{c|}{} \\ 
\hline }
\tabletail{\multicolumn{33}{|r|}{\emph{Continue}} \\ 
\hline}\tablelasttail{\hline}\topcaption{Notes on the stars: \newline
$\S$: stars reported in CTIO as non-variable, which
were used to calibrate the S indexes 
\newline
$\star$: stars with planets  \newline
$\triangleleft$: dMe stars, or stars presumably active \newline
\dag : RS~CVn type \newline
\ddag : BY~Dra type \newline
The spectral types, $V$ magnitudes and $B-V$ colours are from 
\citet{1997A&A...323L..49P}, except the stars indicated with an \1\ which are 
from Simbad database. \newline
The parallexes are from \citet{1997A&A...323L..49P}, 
except the ones marked with a \2\ that are from \citet{1952QB813.J45......}. 
\newline
The gravity and metallicity are from \citet{2001A&A...373..159C}, except the 
ones indicated with a \3\ which are from \citet{1997A&AS..124..299C}.
\newline
The rotational periods are from:
\4\ \citet{1997MNRAS.284..803S},
\5\ \citet{1996ApJ...457L..99B},
\6\ \citet{1984ApJ...279..763N},
\7\ \citet{1990A&A...235..291S},
\8\ \citet{1990ApJS...74..225H} and
\9\ \citet{1989A&A...209..279P}. Also, we estimated periods from \vsini\ using 
\z\ \citet{2000AcA....50..509G} 
and \zz\ \citet{2003A&A...398..647R}. \newline
The last two columns include the slope of the linear regression between 
$F_{\mathrm{H}\alpha}$ and $F_{\mathrm{HK}}$ for the individual spectra of 
each star ($m$, see Fig.~\ref{fig:bins}) and the corresponding correlation 
coefficient ($\rho$). The $F_{\mathrm{H}\alpha}$ which are between 
parentheses were excluded from our analysis in Section ~\ref{sec:FcontHal}, 
since they might be subject to larger errors due to the use of \col\ as a 
proxy for \teff\ for stars with \col $\leq 1.4$.}
\label{tab:stars}

\scriptsize
\begin{supertabular}{ |r @{\hspace{3mm}} l @{\hspace{3mm}} l @{ } l 
@{\hspace{3mm}} l @{\hspace{3mm}} r @{.} l @{\hspace{3mm}} r @{.} l @{ } l 
@{\hspace{3mm}} r @{.} l @{ } l @{\hspace{3mm}} r @{.} l @{\hspace{3mm}} r @{.} 
l @{ } l @{\hspace{3mm}} r @{} r @{.} l @{ } l @{\hspace{3mm}} r r @{.} l  r 
@{.} l r @{.} l r @{.} l r @{.} l| }      

1835 & 17.3 & BE Cet & $\S$ \ddag & G3V & 6 & 390 & 0 & 659 &  & 49 & 05 &  & 4 & 22 & -0 & 01 &  &  & 7 & 78 & \4\ & 14 & 0 & 35 & 187 & 20 & 33 & 75 & 0 & 28 & 0 & 32 \\ 
3405 & 24A &  &  & G0/G1V & 6 & 790 & 0 & 639 &  & 22 & 79 &  & \multicolumn{2}{c}{ }   & \multicolumn{2}{c}{ }   &  &  & \multicolumn{2}{c}{ }   &  & 5 & 0 & 33 & 219 & 10 & 46 & 33 & -3 & 13 & -0 & 88 \\ 
3443 & 25A &  & $\S$ & K1V+... & 5 & 570 & 0 & 715 &  & 64 & 38 &  & 4 & 57 & -0 & 16 & \3\ &  & 30 & 00 & \5\ & 10 & 0 & 17 & 82 & 95 & 29 & 09 & 0 & 76 & 0 & 60 \\ 
3795 & 27.2 &  & $\S$ & G3/G5V & 6 & 140 & 0 & 718 &  & 35 & 02 &  & 3 & 75 & -0 & 70 &  &  & 32 & 00 & \5\ & 10 & 0 & 15 & 70 & 85 & 28 & 05 & 0 & 31 & 0 & 38 \\ 
4308 & 31.5 &  &  & G3V & 6 & 550 & 0 & 655 &  & 45 & 76 &  & 4 & 29 & -0 & 47 &  &  & \multicolumn{2}{c}{ }   &  & 13 & 0 & 16 & 94 & 45 & 30 & 47 & 0 & 04 & 0 & 18 \\ 
4967 & 40 &  &  & K5V & 8 & 950 & 1 & 290 &  & 65 & 81 &  & \multicolumn{2}{c}{ }   & \multicolumn{2}{c}{ }   &  &  & \multicolumn{2}{c}{ }   &  & 1 & 1 & 49 & 33 & 55 & 10 & 66 & \multicolumn{2}{c}{}   & \multicolumn{2}{c|}{}   \\ 
5869 &  &  & & K4V & 7 & 510 & 1 & 473 &  & 4 & 76 &  & \multicolumn{2}{c}{ }
& \multicolumn{2}{c}{ }   &  &  & \multicolumn{2}{c}{ }   &  & 3 & 0 & 18 & 1
& 11 & (31 & 14) & 0 & 03 & 0 & 22 \\ 
9562 &  &  & $\S$ & G2IV & 5 & 750 & 0 & 639 &  & 33 & 71 &  & 3 & 90 & 0 & 18 &  &  & 29 & 00 & \5\ & 9 & 0 & 14 & 84 & 80 & 30 & 39 & 0 & 02 & 0 & 12 \\ 
10700 & 71 & tau Cet & $\S$ & G8V & 3 & 490 & 0 & 727 &  & 274 & 17 &  & 4 & 30 & -0 & 59 &  &  & 34 & 50 & \4\ & 12 & 0 & 18 & 78 & 55 & 28 & 10 & -0 & 02 & -0 & 53 \\ 
11131 &  & chi Cet B & $\S$ & G0 & 6 & 720 & 0 & 654 &  & 43 & 47 &  & 4 & 50 & -0 & 06 &  & $<$ & 5 & 21 & \z\ & 10 & 0 & 33 & 213 & 10 & 36 & 16 & 0 & 11 & 0 & 41 \\ 
12759 &  &  &  & G3V & 7 & 300 & 0 & 694 &  & 20 & 34 &  & \multicolumn{2}{c}{ }   & \multicolumn{2}{c}{ }   &  &  & \multicolumn{2}{c}{ }   &  & 5 & 0 & 42 & 208 & 25 & 35 & 53 & -0 & 50 & -0 & 52 \\ 
13445 &  &  & $\star$ & K0V & 6 & 120 & 0 & 812 &  & 91 & 63 &  & 4 & 75 & -0 & 21 &  &  & 30 & 00 & \4\ & 6 & 0 & 26 & 69 & 80 & 26 & 86 & 0 & 02 & 0 & 16 \\ 
16673 & 3175 &  & $\S$ & F6V & 5 & 790 & 0 & 524 &  & 46 & 42 &  & 4 & 16 & -0 & 01 &  &  & 7 & 00 & \5\ & 10 & 0 & 22 & 225 & 85 & 38 & 10 & 0 & 21 & 0 & 30 \\ 
17051 &  & iot Hor & $\star$ & G3IV & 5 & 400 & 0 & 561 &  & 58 & 00 &  & 4 & 22 & -0 & 04 &  &  & 7 & 90 & \4\ & 6 & 0 & 23 & 193 & 10 & 37 & 36 & 1 & 00 & 0 & 55 \\ 
17576 &  &  &  & G0V & 7 & 830 & 0 & 609 &  & 6 & 13 &  & \multicolumn{2}{c}{ }   & \multicolumn{2}{c}{ }   &  &  & \multicolumn{2}{c}{ }   &  & 5 & 0 & 84 & 614 & 00 & 54 & 48 & -1 & 20 & -0 & 95 \\ 
17925 & 117 & EP Eri & $\S$ \dag & K1V & 6 & 050 & 0 & 862 &  & 96 & 33 &  & 4 & 60 & 0 & 10 &  &  & 6 & 76 & \4\ & 8 & 0 & 65 & 140 & 00 & 31 & 17 & -0 & 14 & -0 & 22 \\ 
19034 & 121.2 &  &  & G5 & 8 & 080 & 0 & 677 &  & 28 & 33 &  & \multicolumn{2}{c}{ }   & \multicolumn{2}{c}{ }   &  &  & \multicolumn{2}{c}{ }   &  & 9 & 0 & 18 & 93 & 70 & 30 & 70 & -0 & 06 & -0 & 20 \\ 
19467 & 3200 &  &  & G3V & 6 & 970 & 0 & 645 &  & 31 & 76 &  & \multicolumn{2}{c}{ }   & \multicolumn{2}{c}{ }   &  &  & \multicolumn{2}{c}{ }   &  & 11 & 0 & 15 & 88 & 15 & 30 & 57 & 0 & 17 & 0 & 28 \\ 
19994 &  & 94 Cet & $\star$ & F8V & 5 & 070 & 0 & 575 &  & 44 & 69 &  & 4 & 10 & 0 & 15 &  &  & 8 & 59 & \z\ & 10 & 0 & 17 & 138 & 25 & 33 & 27 & 1 & 94 & 0 & 78 \\ 
20619 & 135 &  &  & G0 & 7 & 050 & 0 & 655 &  & 40 & 52 &  & 4 & 48 & -0 & 20 &  &  & \multicolumn{2}{c}{ }   &  & 12 & 0 & 18 & 105 & 40 & 29 & 49 & 0 & 19 & 0 & 39 \\ 
20766 & 136 & zet01 Ret &  & G2V & 5 & 530 & 0 & 641 &  & 82 & 51 &  & 4 & 54 & -0 & 22 &  & $<$ & 12 & 10 & \zz\ & 13 & 0 & 25 & 157 & 45 & 33 & 10 & 0 & 15 & 0 & 31 \\ 
22049 & 144 & eps Eri & $\S$ \ddag & K2V & 3 & 720 & 0 & 881 &  & 310 & 75 &  & 4 & 38 & -0 & 14 &  &  & 11 & 68 & \4\ & 14 & 0 & 47 & 89 & 90 & 24 & 74 & 0 & 12 & 0 & 42 \\ 
23249 & 150 & del Eri & $\S$ \ddag & K0IV & 3 & 520 & 0 & 915 &  & 110 & 58 &  & 3 & 95 & 0 & 05 &  &  & 71 & 00 & \5\ & 14 & 0 & 13 & 22 & 68 & 19 & 26 & 0 & 07 & 0 & 53 \\ 
25069 &  &  &  & G9V & 5 & 850 & 1 & 001 &  & 22 & 88 &  & \multicolumn{2}{c}{ }   & \multicolumn{2}{c}{ }   &  &  & \multicolumn{2}{c}{ }   &  & 6 & 0 & 12 & 13 & 01 & 15 & 77 & 1 & 30 & 0 & 75 \\ 
26965 & 166A & DY Eri & $\S$ & K1V & 4 & 430 & 0 & 820 &  & 198 & 24 &  & 4 & 31 & -0 & 25 &  &  & 37 & 10 & \4\ & 10 & 0 & 19 & 51 & 55 & 26 & 41 & -0 & 02 & -0 & 15 \\ 
27442 &  & eps Ret & $\star$ & K2IV & 4 & 440 & 1 & 078 &  & 54 & 84 &  & 3 & 30 & 0 & 22 &  &  & 54 & 96 & \zz\ & 10 & 0 & 13 & 10 & 34 & 12 & 42 & 0 & 01 & 0 & 36 \\ 
28246 &  &  &  & F6V & 6 & 380 & 0 & 457 &  & 27 & 00 &  & \multicolumn{2}{c}{ }   & \multicolumn{2}{c}{ }   &  &  & \multicolumn{2}{c}{ }   &  & 6 & 0 & 21 & 303 & 10 & 39 & 55 & 0 & 08 & 0 & 76 \\ 
30495 & 177 & 58 Eri & $\S$ & G3V & 5 & 490 & 0 & 632 &  & 75 & 10 &  & 4 & 30 & -0 & 13 &  &  & 9 & 10 & \4\ & 12 & 0 & 30 & 194 & 40 & 33 & 28 & 0 & 22 & 0 & 47 \\ 
32147 &  &  &  & K3V & 6 & 220 & 1 & 049 &  & 113 & 46 &  & 4 & 57 & 0 & 34 &  &  & 47 & 40 & \4\ & 11 & 0 & 30 & 24 & 55 & 16 & 19 & 0 & 15 & 0 & 68 \\ 
35850 &  &  &  & F7V: & 6 & 300 & 0 & 553 &  & 37 & 26 &  & 4 & 40 & 0 & 00 &  &  & 0 & 91 & \z\ & 6 & 0 & 50 & 482 & 30 & 56 & 47 & 0 & 24 & 0 & 68 \\ 
36395 & 205 &  & & M1V & 7 & 970 & 1 & 474 &  & 175 & 72 &  & 4 & 80 & 0 & 60
& \3\ & $<$ & 7 & 85 & \z\ & 9 & 2 & 44 & 20 & 12 & (6 & 70) & 0 & 33 & 0 & 97 \\ 
36435 &  &  &  & G5V & 6 & 990 & 0 & 755 &  & 51 & 10 &  & 4 & 49 & -0 & 02 &  &  & 11 & 20 & \4\ & 5 & 0 & 44 & 149 & 40 & 28 & 75 & 0 & 04 & 0 & 24 \\ 
37572 &  & UY Pic &  & K0V & 7 & 950 & 0 & 845 &  & 41 & 90 &  & \multicolumn{2}{c}{ }   & \multicolumn{2}{c}{ }   &  &  & \multicolumn{2}{c}{ }   &  & 12 & 0 & 79 & 211 & 05 & 36 & 38 & 0 & 14 & 0 & 34 \\ 
38392 & 216B & gam Lep B & $\S$ & K2V & 6 & 150 & 0 & 940 & \1\ & \multicolumn{2}{c}{ }   &  & 4 & 50 & 0 & 02 &  &  & 17 & 30 & \4\ & 12 & 0 & 48 & 62 & 95 & 22 & 14 & 0 & 08 & 0 & 29 \\ 
38393 & 216A & gam Lep & $\S$ & F7V & 3 & 590 & 0 & 481 &  & 111 & 49 &  & 4 & 10 & -0 & 12 &  &  & 13 & 22 & \z\ & 13 & 0 & 16 & 188 & 15 & 39 & 16 & 0 & 04 & 0 & 06 \\ 
38858 & 1085 &  &  & G4V & 5 & 970 & 0 & 639 &  & 64 & 25 &  & \multicolumn{2}{c}{ }   & \multicolumn{2}{c}{ }   &  &  & \multicolumn{2}{c}{ }   &  & 6 & 0 & 16 & 102 & 85 & 32 & 39 & 0 & 16 & 0 & 34 \\ 
39917 &  & SZ Pic & \dag & G8V & 7 & 890 & 0 & 813 &  & 5 & 13 &  & \multicolumn{2}{c}{ }   & \multicolumn{2}{c}{ }   &  &  & 4 & 48 & \z\ & 10 & 0 & 54 & 158 & 10 & 47 & 64 & -0 & 59 & -0 & 57 \\ 
41824 &  &  & \dag & G6V & 6 & 600 & 0 & 712 &  & 33 & 64 &  & \multicolumn{2}{c}{ }   & \multicolumn{2}{c}{ }   &  &  & \multicolumn{2}{c}{ }   &  & 12 & 0 & 57 & 260 & 30 & 44 & 30 & 0 & 43 & 0 & 86 \\ 
42581 & 229A &  & $\triangleleft$ & M1/M2V & 8 & 150 & 1 & 487 &  & 173 & 19 &
& \multicolumn{2}{c}{ }   & \multicolumn{2}{c}{ }   &  &  & 7 & 74 & \z\ & 9 &
1 & 66 & 9 & 72 & (4 & 73) & 0 & 15 & 0 & 61 \\ 
43587 & 231.1A &  &  & G0.5Vb & 5 & 700 & 0 & 610 &  & 51 & 76 &  & \multicolumn{2}{c}{ }   & -0 & 08 &  &  & 20 & 00 & \5\ & 8 & 0 & 15 & 100 & 40 & 32 & 81 & -0 & 13 & -0 & 65 \\ 
45067 &  &  & $\S$ & F8V & 5 & 880 & 0 & 564 &  & 30 & 22 &  & 4 & 17 & -0 & 16 &  &  & 8 & 00 & \5\ & 11 & 0 & 14 & 125 & 20 & 32 & 46 & 0 & 05 & 0 & 27 \\ 
45270 &  &  &  & G1V & 6 & 530 & 0 & 614 &  & 42 & 56 &  & \multicolumn{2}{c}{ }   & \multicolumn{2}{c}{ }   &  &  & \multicolumn{2}{c}{ }   &  & 6 & 0 & 39 & 268 & 25 & 38 & 43 & 0 & 45 & 0 & 70 \\ 
48189 & 3400A &  &  & G1/G2V & 6 & 150 & 0 & 624 &  & 46 & 15 &  & \multicolumn{2}{c}{ }   & \multicolumn{2}{c}{ }   &  &  & \multicolumn{2}{c}{ }   &  & 7 & 0 & 44 & 274 & 35 & 45 & 52 & 0 & 57 & 0 & 62 \\ 
52265 &  &  & $\star$ & G0III-IV & 6 & 290 & 0 & 572 &  & 35 & 63 &  & 4 & 29 & 0 & 21 &  &  & 9 & 42 & \z\ & 3 & 0 & 16 & 130 & 10 & 36 & 05 & 2 & 24 & 1 & 00 \\ 
54579 &  & V361 Pup &  & G0V & 8 & 030 & 0 & 606 &  & 16 & 95 &  & \multicolumn{2}{c}{ }   & \multicolumn{2}{c}{ }   &  &  & \multicolumn{2}{c}{ }   &  & 3 & 0 & 47 & 337 & 30 & 59 & 99 & 0 & 30 & 0 & 87 \\ 
59967 & 3446 &  &  & G3V & 6 & 660 & 0 & 641 &  & 45 & 93 &  & \multicolumn{2}{c}{ }   & \multicolumn{2}{c}{ }   &  &  & 10 & 26 & \z\ & 5 & 0 & 38 & 232 & 15 & 36 & 44 & 0 & 18 & 0 & 38 \\ 
75289 &  &  & $\star$ & G0Ia0: & 6 & 350 & 0 & 578 &  & 34 & 55 &  & 4 & 51 & 0 & 28 &  &  & \multicolumn{2}{c}{ }   &  & 8 & 0 & 15 & 118 & 45 & 33 & 25 & 0 & 05 & 0 & 84 \\ 
82106 & 349 &  & $\triangleleft$ & K3V & 7 & 200 & 1 & 002 &  & 78 & 87 &  & \multicolumn{2}{c}{ }   & \multicolumn{2}{c}{ }   &  &  & 13 & 30 & \4\ & 15 & 0 & 32 & 32 & 57 & 17 & 93 & 0 & 18 & 0 & 45 \\ 
94683 &  &  & & K4III & 5 & 950 & 1 & 736 &  & 1 & 96 &  & \multicolumn{2}{c}{
}   & \multicolumn{2}{c}{ }   &  &  & \multicolumn{2}{c}{ }   &  & 7 & 0 & 37
& 0 & 10 & (0 & 26) & -0 & 36 & -0 & 69 \\ 
97233 & 416 &  &  & K4V & 9 & 060 & 1 & 203 &  & 44 & 00 &  & \multicolumn{2}{c}{ }   & \multicolumn{2}{c}{ }   &  &  & 7 & 60 & \6\ & 10 & 0 & 70 & 21 & 62 & 13 & 71 & -0 & 00 & -0 & 00 \\ 
101581 & 435 &  & $\triangleleft$ & K5V & 7 & 770 & 1 & 064 &  & 79 & 91 &  & \multicolumn{2}{c}{ }   & \multicolumn{2}{c}{ }   &  &  & \multicolumn{2}{c}{ }   &  & 9 & 0 & 49 & 29 & 16 & 17 & 98 & 0 & 06 & 0 & 41 \\ 
103112 & 3689A &  &  & K0 & 7 & 610 & 1 & 055 &  & 12 & 39 &  & \multicolumn{2}{c}{ }   & \multicolumn{2}{c}{ }   &  &  & \multicolumn{2}{c}{ }   &  & 9 & 0 & 15 & 13 & 17 & 14 & 55 & -0 & 01 & -0 & 24 \\ 
105115 &  &  &  & K2/K3V & 6 & 910 & 1 & 391 &  & 2 & 85 &  & \multicolumn{2}{c}{ }   & \multicolumn{2}{c}{ }   &  &  & \multicolumn{2}{c}{ }   &  & 5 & 0 & 15 & 1 & 49 & 3 & 77 & 0 & 03 & 0 & 76 \\ 
114762 &  &  & $\star$ & F9V & 7 & 300 & 0 & 525 &  & 24 & 65 &  & 4 & 10 & -0 & 82 &  &  & 52 & 16 & \z\ & 3 & 0 & 16 & 145 & 95 & 39 & 76 & \multicolumn{2}{c}{}   & \multicolumn{2}{c|}{}   \\ 
117176 &  & 70 Vir & $\star$ & G5V & 4 & 970 & 0 & 714 &  & 55 & 22 &  & 3 & 83 & -0 & 11 &  &  & \multicolumn{2}{c}{ }   &  & 4 & 0 & 10 & 38 & 85 & 28 & 15 & -0 & 00 & -0 & 06 \\ 
118100 & 517 & EQ Vir & $\triangleleft$ & K7V & 9 & 240 & 1 & 210 &  & 50 & 54 &  & 4 & 50 & -0 & 15 & \3\ &  & 3 & 90 & \4\ & 9 & 3 & 68 & 118 & 55 & 25 & 63 & 0 & 75 & 0 & 98 \\ 
119285 &  & V851 Cen & \dag & K1Vp & 7 & 640 & 1 & 068 &  & 13 & 13 &  & \multicolumn{2}{c}{ }   & \multicolumn{2}{c}{ }   &  &  & 12 & 03 & \7\ & 5 & 0 & 51 & 34 & 76 & 21 & 57 & 0 & 39 & 1 & 00 \\ 
119850 & 526 &  & $\triangleleft$ & M3V & 8 & 460 & 1 & 435 &  & 184 & 13 &  &
\multicolumn{2}{c}{ }   & \multicolumn{2}{c}{ }   &  &  & 8 & 60 & \z\ & 9 & 0
& 57 & 5 & 45 & (5 & 81) & 0 & 40 & 0 & 53 \\ 
120136 &  & tau Boo & $\star$ & F7V & 4 & 500 & 0 & 508 &  & 64 & 12 &  & 4 & 18 & 0 & 32 &  &  & 4 & 00 & \5\ & 3 & 0 & 19 & 203 & 95 & 36 & 65 & -0 & 14 & -0 & 20 \\ 
122303 & 536 &  & & K5 & 9 & 710 & 1 & 461 &  & 98 & 26 &  &
\multicolumn{2}{c}{ }   & \multicolumn{2}{c}{ }   &  &  & \multicolumn{2}{c}{
}   &  & 8 & 0 & 97 & 8 & 11 & (6 & 99) & -0 & 04 & -0 & 05 \\ 
123732 &  & V759 Cen &  & G0V & 7 & 540 & 0 & 594 &  & 15 & 88 &  & \multicolumn{2}{c}{ }   & \multicolumn{2}{c}{ }   &  &  & \multicolumn{2}{c}{ }   &  & 2 & 0 & 44 & 334 & 85 & 62 & 73 & 0 & 56 & \multicolumn{2}{c|}{}   \\ 
125072 & 542 &  &  & K3V & 6 & 660 & 1 & 017 &  & 84 & 50 &  & 4 & 50 & 0 & 26 &  &  & \multicolumn{2}{c}{ }   &  & 10 & 0 & 27 & 20 & 42 & 17 & 49 & -0 & 09 & -0 & 42 \\ 
128620 & 599A & alf Cen A &  & G2V & -0 & 010 & 0 & 710 &  & 742 & 12 &  & 4 & 31 & 0 & 22 &  &  & 29 & 00 & \4\ & 7 & 0 & 15 & 71 & 60 & 27 & 31 & 0 & 00 & 0 & 02 \\ 
128621 &  & alf Cen B &  & K1V & 1 & 350 & 0 & 900 &  & 742 & 12 &  & 4 & 51 & 0 & 24 &  &  & 42 & 00 & \4\ & 9 & 0 & 20 & 35 & 56 & 21 & 55 & 0 & 30 & 0 & 85 \\ 
130948 &  &  &  & G2V & 5 & 860 & 0 & 576 &  & 55 & 73 &  & 4 & 18 & -0 & 20 &  &  & 6 & 40 & \z\ & 6 & 0 & 30 & 205 & 70 & 37 & 34 & -0 & 00 & -0 & 02 \\ 
131977 &  &  &  & K4V & 5 & 720 & 1 & 024 &  & 169 & 32 &  & 4 & 58 & 0 & 03 &  &  & 44 & 60 & \4\ & 9 & 0 & 51 & 33 & 00 & 19 & 37 & 0 & 01 & 0 & 11 \\ 
144253 & 610 &  &  & K3/K4V & 7 & 390 & 1 & 043 &  & 53 & 93 &  & \multicolumn{2}{c}{ }   & -0 & 03 &  &  & \multicolumn{2}{c}{ }   &  & 10 & 0 & 24 & 19 & 72 & 19 & 38 & -0 & 21 & -0 & 44 \\ 
146233 & 616 & 18 Sco &  & G1V & 5 & 490 & 0 & 652 &  & 71 & 30 &  & 4 & 49 & 0 & 05 &  &  & 23 & 70 & \4\ & 15 & 0 & 17 & 95 & 00 & 31 & 49 & 0 & 03 & 0 & 06 \\ 
147513 & 620.1A &  &  & G3/G5V & 5 & 370 & 0 & 625 &  & 77 & 69 &  & 4 & 52 & 0 & 03 &  &  & 8 & 50 & \4\ & 11 & 0 & 32 & 193 & 30 & 34 & 96 & 0 & 15 & 0 & 54 \\ 
150433 & 634.1 &  &  & G0 & 7 & 210 & 0 & 631 &  & 33 & 84 &  & \multicolumn{2}{c}{ }   & \multicolumn{2}{c}{ }   &  &  & \multicolumn{2}{c}{ }   &  & 8 & 0 & 17 & 104 & 85 & 34 & 12 & 0 & 17 & 0 & 97 \\ 
151770 &  &  &  & G3/G5V & 8 & 340 & 0 & 676 &  & 7 & 45 &  & \multicolumn{2}{c}{ }   & \multicolumn{2}{c}{ }   &  &  & \multicolumn{2}{c}{ }   &  & 4 & 0 & 62 & 313 & 40 & 61 & 08 & -0 & 64 & -0 & 97 \\ 
152391 & 641 & V2292 Oph & $\S$ \ddag & G8V & 6 & 650 & 0 & 749 &  & 59 & 04 &  & \multicolumn{2}{c}{ }   & \multicolumn{2}{c}{ }   &  &  & 11 & 43 & \4\ & 12 & 0 & 41 & 147 & 10 & 32 & 02 & 0 & 03 & 0 & 17 \\ 
156026 & 664 & 36 Oph C & \dag & K5V & 6 & 330 & 1 & 144 &  & 167 & 56 &  & 4 & 67 & -0 & 21 &  &  & 18 & 50 & \4\ & 7 & 0 & 76 & 29 & 45 & 15 & 07 & -0 & 06 & -0 & 40 \\ 
156425 &  &  &  & K4V: & 7 & 810 & 1 & 098 &  & \multicolumn{2}{c}{ }   &  & \multicolumn{2}{c}{ }   & -0 & 20 &  &  & \multicolumn{2}{c}{ }   &  & 7 & 0 & 41 & 5 & 79 & 13 & 33 & 0 & 02 & 0 & 06 \\ 
157881 & 673 &  &  & K7V & 7 & 540 & 1 & 359 &  & 129 & 54 &  & 4 & 70 & -0 & 20 &  &  & 7 & 29 & \z\ & 10 & 0 & 78 & 8 & 44 & 8 & 12 & 0 & 05 & 0 & 71 \\ 
158614 & 678A &  & $\S$ & G8IV-V & 5 & 310 & 0 & 715 &  & 60 & 80 &  & 4 & 04 & -0 & 05 &  &  & 34 & 00 & \5\ & 12 & 0 & 22 & 66 & 25 & 27 & 97 & -0 & 10 & -0 & 58 \\ 
160691 & 691 & mu Ara &  & G5V & 5 & 120 & 0 & 694 &  & 65 & 46 &  & 4 & 20 & 0 & 20 &  &  & 14 & 04 & \zz\ & 8 & 0 & 14 & 63 & 50 & 28 & 82 & -0 & 00 & -0 & 02 \\ 
165185 &  &  &  & G3V & 5 & 940 & 0 & 615 &  & 57 & 58 &  & 4 & 49 & -0 & 06 &  &  & 5 & 90 & \4\ & 10 & 0 & 30 & 197 & 75 & 36 & 55 & 0 & 23 & 0 & 91 \\ 
172051 &  &  &  & G5V & 5 & 850 & 0 & 673 &  & 77 & 02 &  & \multicolumn{2}{c}{ }   & \multicolumn{2}{c}{ }   &  &  & \multicolumn{2}{c}{ }   &  & 13 & 0 & 17 & 92 & 60 & 31 & 32 & -0 & 17 & -0 & 42 \\ 
173560 & 725.3 &  &  & G3V & 8 & 690 & 0 & 649 &  & 11 & 65 &  & \multicolumn{2}{c}{ }   & \multicolumn{2}{c}{ }   &  &  & \multicolumn{2}{c}{ }   &  & 11 & 0 & 19 & 109 & 75 & 33 & 41 & 0 & 03 & 0 & 87 \\ 
177996 & 4096 &  &  & K1V & 7 & 890 & 0 & 862 &  & 31 & 48 &  & \multicolumn{2}{c}{ }   & \multicolumn{2}{c}{ }   &  &  & \multicolumn{2}{c}{ }   &  & 9 & 0 & 64 & 133 & 05 & 34 & 16 & 0 & 01 & 0 & 03 \\ 
180617 & 752A &  & $\triangleleft$ & M3.5V & 9 & 120 & 1 & 464 &  & 170 & 26 &
& \multicolumn{2}{c}{ }   & \multicolumn{2}{c}{ }   &  &  & 16 & 80 & \z\ & 8
& 1 & 08 & 10 & 03 & (7 & 80) & 0 & 04 & 0 & 20 \\ 
187923 & 4126 &  &  & G0V & 6 & 160 & 0 & 642 &  & 36 & 15 &  & 3 & 97 & -0 & 20 &  & $<$ & 3 & 72 & \z\ & 10 & 0 & 14 & 85 & 00 & 35 & 15 & 0 & 02 & 0 & 20 \\ 
188088 & 770 & V4200 Sgr & \ddag & K3/K4V & 6 & 220 & 1 & 017 &  & 70 & 34 &  & \multicolumn{2}{c}{ }   & \multicolumn{2}{c}{ }   &  &  & 16 & 50 & \8\ & 6 & 0 & 59 & 56 & 35 & 19 & 15 & -0 & 46 & -0 & 58 \\ 
189567 & 776 &  &  & G2V & 6 & 070 & 0 & 648 &  & 56 & 45 &  & 4 & 10 & -0 & 30 &  &  & \multicolumn{2}{c}{ }   &  & 8 & 0 & 17 & 94 & 00 & 32 & 60 & 0 & 02 & 0 & 28 \\ 
197076 & 797A &  &  & G5V & 6 & 430 & 0 & 611 &  & 47 & 65 &  & \multicolumn{2}{c}{ }   & \multicolumn{2}{c}{ }   &  &  & 13 & 21 & \z\ & 7 & 0 & 16 & 90 & 60 & 34 & 66 & -0 & 04 & -0 & 15 \\ 
197214 & 4157 &  &  & G3/G5V & 6 & 950 & 0 & 671 &  & 44 & 57 &  & \multicolumn{2}{c}{ }   & \multicolumn{2}{c}{ }   &  &  & \multicolumn{2}{c}{ }   &  & 12 & 0 & 19 & 108 & 05 & 30 & 37 & -0 & 00 & -0 & 04 \\ 
202628 & 825.2 &  &  & G5V & 6 & 750 & 0 & 637 &  & 42 & 04 &  & 4 & 52 & -0 & 14 &  &  & \multicolumn{2}{c}{ }   &  & 10 & 0 & 24 & 158 & 00 & 32 & 41 & -0 & 01 & -0 & 05 \\ 
202917 &  &  &  & G5V & 8 & 650 & 0 & 690 &  & 21 & 81 &  & \multicolumn{2}{c}{ }   & \multicolumn{2}{c}{ }   &  &  & \multicolumn{2}{c}{ }   &  & 6 & 0 & 68 & 358 & 35 & 51 & 16 & 0 & 34 & 0 & 66 \\ 
202996 &  &  &  & G0V & 7 & 460 & 0 & 614 &  & 18 & 76 &  & \multicolumn{2}{c}{ }   & \multicolumn{2}{c}{ }   &  &  & \multicolumn{2}{c}{ }   &  & 2 & 0 & 26 & 203 & 65 & 38 & 26 & 0 & 42 & \multicolumn{2}{c|}{}   \\ 
203019 &  &  &  & G5V & 7 & 840 & 0 & 687 &  & 27 & 49 &  & \multicolumn{2}{c}{ }   & \multicolumn{2}{c}{ }   &  &  & \multicolumn{2}{c}{ }   &  & 4 & 0 & 49 & 253 & 25 & 34 & 40 & 0 & 09 & 0 & 89 \\ 
203244 &  &  &  & G5V & 6 & 980 & 0 & 723 &  & 48 & 86 &  & 4 & 49 & -0 & 21 &  &  & \multicolumn{2}{c}{ }   &  & 1 & 0 & 40 & 183 & 30 & 33 & 63 & \multicolumn{2}{c}{}   & \multicolumn{2}{c|}{}   \\ 
209100 & 845 & eps Ind & $\triangleleft$ & K5V & 4 & 690 & 1 & 056 &  & 275 & 76 &  & 4 & 50 & -0 & 23 &  &  & 22 & 00 & \4\ & 11 & 0 & 44 & 31 & 67 & 16 & 02 & -0 & 14 & -0 & 51 \\ 
210918 & 851.2 &  &  & G5V & 6 & 230 & 0 & 648 &  & 45 & 19 &  & 4 & 43 & -0 & 18 &  &  & \multicolumn{2}{c}{ }   &  & 12 & 0 & 16 & 96 & 35 & 29 & 92 & 0 & 03 & 0 & 10 \\ 
212330 & 857 &  &  & F9V & 5 & 310 & 0 & 665 &  & 48 & 81 &  & 4 & 00 & -0 & 04 &  & $<$ & 21 & 05 & \z\ & 10 & 0 & 14 & 72 & 75 & 29 & 99 & 0 & 01 & 0 & 07 \\ 
213240 &  &  & $\star$ & G4IV & 6 & 810 & 0 & 603 &  & 24 & 54 &  & \multicolumn{2}{c}{ }   & \multicolumn{2}{c}{ }   &  &  & \multicolumn{2}{c}{ }   &  & 3 & 0 & 18 & 132 & 65 & 33 & 91 & 0 & 09 & 0 & 96 \\ 
213941 & 863.3 &  &  & G5V & 7 & 580 & 0 & 670 &  & 30 & 98 &  & \multicolumn{2}{c}{ }   & -0 & 42 &  &  & \multicolumn{2}{c}{ }   &  & 9 & 0 & 19 & 124 & 10 & 29 & 15 & -0 & 16 & -0 & 41 \\ 
215768 &  &  &  & G0V & 7 & 490 & 0 & 589 &  & 25 & 34 &  & \multicolumn{2}{c}{ }   & \multicolumn{2}{c}{ }   &  &  & \multicolumn{2}{c}{ }   &  & 5 & 0 & 31 & 233 & 15 & 37 & 36 & 0 & 17 & 0 & 32 \\ 
216803 &  & TW PsA &  & K4Vp & 6 & 480 & 1 & 094 &  & 130 & 94 &  & \multicolumn{2}{c}{ }   & \multicolumn{2}{c}{ }   &  &  & 10 & 30 & \4\ & 3 & 1 & 00 & 61 & 20 & 18 & 68 & -0 & 75 & -0 & 93 \\ 
217343 &  &  &  & G3V & 7 & 470 & 0 & 655 &  & 31 & 22 &  & \multicolumn{2}{c}{ }   & \multicolumn{2}{c}{ }   &  &  & \multicolumn{2}{c}{ }   &  & 2 & 0 & 47 & 285 & 55 & 36 & 22 & -0 & 03 & \multicolumn{2}{c|}{}   \\ 
219709 &  &  &  & G2V & 7 & 500 & 0 & 632 &  & 25 & 33 &  & \multicolumn{2}{c}{ }   & -0 & 02 &  &  & \multicolumn{2}{c}{ }   &  & 8 & 0 & 22 & 147 & 80 & 31 & 15 & 0 & 08 & 0 & 24 \\ 
219834 & 894.2A & 94 Aqr & $\S$ & G6/G8IV & 5 & 200 & 0 & 787 &  & 48 & 22 & \2\ & 3 & 90 & 0 & 09 &  &  & 42 & 50 & \5\ & 12 & 0 & 16 & 53 & 55 & 24 & 71 & -0 & 22 & -0 & 51 \\ 
225213 & 1 &  & & M2V & 8 & 560 & 1 & 462 &  & 229 & 33 &  &
\multicolumn{2}{c}{ }   & \multicolumn{2}{c}{ }   &  &  & \multicolumn{2}{c}{
}   &  & 10 & 0 & 36 & 3 & 06 & (7 & 53) & 0 & 03 & 0 & 10 \\ 
 & 375 & LU Vel & $\triangleleft$ & M5 & 11 & 270 & 1 & 555 &  & 62 & 88 &  &
\multicolumn{2}{c}{ }   & \multicolumn{2}{c}{ }   &  &  & \multicolumn{2}{c}{
}   &  & 14 & 2 & 56 & 6 & 33 & (4 & 74) & 0 & 34 & 0 & 46 \\ 
 & 388 & AD Leo & $\triangleleft$ & M3.5Ve & 9 & 430 & 1 & 540 & \1\ & 213 &
00 & \2\ & \multicolumn{2}{c}{ }   & \multicolumn{2}{c}{ }   &  &  & 2 & 60 &
\9\ & 7 & 9 & 04 & 30 & 89 & (13 & 15) & 0 & 15 & 0 & 38 \\ 
 & 479 &  & & M3 & 10 & 650 & 1 & 470 &  & 103 & 54 &  & \multicolumn{2}{c}{ }
& \multicolumn{2}{c}{ }   &  &  & \multicolumn{2}{c}{ }   &  & 7 & 1 & 78 & 13
& 32 & (10 & 16) & 0 & 07 & 0 & 41 \\ 
 & 551 & Proxima Cen & $\triangleleft$ & M5Ve & 11 & 010 & 1 & 807 &  & 772 &
33 &  & \multicolumn{2}{c}{ }   & \multicolumn{2}{c}{ }   &  &  &
\multicolumn{2}{c}{ }   &  & 22 & 4 & 10 & 1 & 68 & (2 & 05) & 0 & 10 & 0 & 31 \\ 
 & 699 & Barnard's star & $\triangleleft$ & sdM4 & 9 & 540 & 1 & 570 &  & 549
& 01 &  & \multicolumn{2}{c}{ }   & \multicolumn{2}{c}{ }   &  &  & 2 & 09 &
\z\ & 11 & 0 & 93 & 2 & 61 & (2 & 48) & 0 & 04 & 0 & 20 \\ 
 & 729 & V1216 Sgr & $\triangleleft$ & M3.5Ve & 10 & 370 & 1 & 510 &  & 336 &
48 &  & \multicolumn{2}{c}{ }   & \multicolumn{2}{c}{ }   &  &  &
\multicolumn{2}{c}{ }   &  & 1 & 6 & 98 & 79 & 85 & (30 & 12) & \multicolumn{2}{c}{}   & \multicolumn{2}{c|}{}   \\ 

\hline
\end{supertabular} 
\end{center}
 
\end{document}